\documentclass[12pt]{article}
\usepackage{amsmath,amsfonts,amsthm}
\usepackage{epsf}
\newcommand{\be}{\begin{equation}}
\newcommand{\ee}{\end{equation}}
\newcommand{\ba}{\begin{array}}
\newcommand{\ea}{\end{array}}
\newcommand{\bea}{\begin{eqnarray}}
\newcommand{\eea}{\end{eqnarray}}
\newcommand{\beaa}{\begin{eqnarray*}}
\newcommand{\eeaa}{\end{eqnarray*}}
\newcommand{\Ga}{\Gamma}
\newcommand{\la}{\lambda}
\newcommand{\si}{\sigma}
\renewcommand{\th}{\theta}
\newcommand{\rb}{\right]}
\newcommand{\lb}{\left[}
\newcommand{\bl}{\biggl(}
\newcommand{\br}{\biggr)}
\renewcommand{\(}{\left(}
\renewcommand{\)}{\right)}
\newcommand{\ao}{a_n^{(1)}}
\newcommand{\ato}{a_2^{(1)}}
\newcommand{\nn}{\nonumber}
\newcommand{\fns}{\footnotesize}
\newcommand{\scs}{\scriptsize}
\newcommand{\shs}{\shortstack}
\newcommand{\ol}{\overline}
\newcommand{\noi}{\noindent}
\newcommand{\hs}{\hspace}
\newcommand{\vs}{\vspace}
\newcommand{\lra}{\longrightarrow}
\advance\textheight by \topskip
\makeatletter

\@addtoreset{equation}{section}
\def\section{\@startsection {section}{1}{\z@}{-8.5ex plus -1ex minus
 -.2ex}{3.3ex plus .2ex}{\large\bf\centering}}
\def\subsection{\@startsection{subsection}{2}{\z@}{-3.25ex plus
 -1ex minus -.2ex}{1.5ex plus .2ex}{\bf}}
\def\subsubsection{\@startsection{subsubsection}{3}{\z@}{-3.25ex plus%
 -1ex minus -.2ex}{1.5ex plus .2ex}{\sl}}
\newcommand{\bri}{\br_{\hs{-5pt} I}}

\newcommand{\acknowledgements}{\vspace{1cm}\noindent{\bf\large{Acknowledgements}}}
\newcommand{\eq}[1]{eq.(\ref{#1})}
\newcommand{\bphi}{{\boldsymbol \phi}}
\newcommand{\weight}{{\boldsymbol \lambda}}
\newcommand{\aroot}{{\boldsymbol \alpha}}

\newcommand{\bp}{C}    
\newcommand{\e}{{\boldsymbol{e}}}        
\newcommand{\er}{\boldsymbol{\epsilon}} 
\newcommand{\kextra}{M} 
\theoremstyle{plain}
\newtheorem{theorem}{Theorem}[section]
\newtheorem{proposition}[theorem]{Proposition}

\newtheorem{corollary}[theorem]{Corollary}
\theoremstyle{definition}

\begin{document}
\begin{titlepage}
\vspace*{-2cm}

\begin{flushright}
KCL-MTH-99-10\\
DTP--99/23\\
hep-th/9904002 \\
\end{flushright}
\vspace{0.3cm}
\begin{center}
{\Large
{\bf Particle Reflection Amplitudes\\
\vspace{2mm}
in $\ao$ Toda Field Theories}}\\
\vspace{1cm}
{\large \bf G.\ W.\ Delius\footnote{\noi E-mail:
{\tt delius@mth.kcl.ac.uk},~~~~home page:
{\tt http://www.mth.kcl.ac.uk/\~{}delius/}} and
G.\ M.\ Gandenberger}\footnote{\noi E-mail:
{\tt G.M.Gandenberger@durham.ac.uk}}\\
\vs{0.3cm}
{${}^a$\em Department of Mathematics\\
King's College London\\
Strand, London WC2R 2LS, U.K.}\\
\vs{0.3cm}
{${}^b$\em Department of Mathematical Sciences\\
Durham University\\
Durham DH1 3LE, U.K.}\\
\vspace{1cm}
{\bf{ABSTRACT}}
\end{center}
\begin{quote}
We determine the exact quantum particle reflection amplitudes
for all known vacua of $a_n^{(1)}$ affine Toda theories
on the half-line with
integrable boundary conditions.
(Real non-singular vacuum solutions are known for about half
of all the classically integrable boundary conditions.)
To be able to do this we use the fact
that the particles can be identified with the
analytically continued breather solutions, and that the real vacuum
solutions are obtained by analytically continuing stationary
soliton solutions.
We thus obtain the particle reflection amplitudes
from the corresponding breather reflection amplitudes. These
in turn we calculate by bootstrapping from soliton reflection
matrices which we obtained as solutions of the boundary
Yang-Baxter equation (reflection equation).

We study the pole structure of the
particle reflection amplitudes and uncover an
unexpectedly rich spectrum of excited boundary states, created by
particles binding to the boundary. For $a_2^{(1)}$ and $a_4^{(1)}$
Toda theories we calculate the reflection amplitudes
for particle reflection off all these excited boundary states.
We are able to explain all physical strip poles in these
reflection factors either in terms of boundary bound states or a
generalisation of the Coleman-Thun mechanism.

\end{quote}

\vfill

\end{titlepage}

\section{Introduction and Overview}
The study of two-dimensional integrable quantum field theories is
interesting because it allows one to calculate  exactly things like
the quantum behaviour of solitons,
the weak--strong coupling duality,
the effects of space-time boundaries, etc. Exact
studies in two dimensions allow one to speculate more confidently
about these phenomena also in higher dimensions.
A particularly rewarding class of integrable quantum field
theories to study are the affine Toda field theories (ATFTs). Some
of their nice features include: they
exhibit weak--strong coupling duality,
they have quantum group symmetries \cite{ber},
they have solitons and breathers \cite{hol,oli}, and,
most importantly for this paper,
they have a rich set of integrable boundary conditions
\cite{corri94,bowco95}.

For the particles of real coupling affine Toda theories without
boundaries it
has been possible to find the exact S-matrices, describing the
evolution of arbitrary asymptotic incoming particle states into
asymptotic outgoing particle states \cite{Arin,brade90,chri,deliu92}.
It has been a longstanding challenge to
extend these results to the half-line with integrable boundary
conditions.

We will only deal with $a_n^{(1)}$ Toda theory in this paper.
The classical equation of motion for the $n$-component bosonic
field of $\ao$ Toda theory is
\begin{equation}
\partial_t^2{\bphi}-\partial_x^2{\bphi}+\frac{m^2}{\beta}\sum_{i= 0}^n
\aroot_ie^{\beta\aroot_i\cdot \bphi}= 0.
\label{phieom}
\end{equation}
The $\alpha_i$, $i=1,\cdots,n$ are the simple roots of the Lie algebra
$a_n=\text{sl}_{n+1}$ and $\alpha_0$ is minus the highest root.
We will use units so that the mass scale $m= 1$.
Note that the coupling constant
$\beta$ could be removed from the
equations of motion by rescaling the field and
therefore the coupling
constant plays a role only in the quantum theory.

It has been discovered in \cite{corri94,bowco95} that
this equation of motion
can be restricted to the left half-line $x<0$ without
losing integrability if one imposes a boundary condition at
$x= 0$ of the form
\begin{equation}\label{bcond}
\left.\beta\,\partial_x\bphi+\sum_{i= 0}^n \bp_i\,\aroot_i
e^{\beta\aroot_i\cdot\bphi/2}\right|_{x= 0}= 0,
\end{equation}
where the boundary parameters $C_i$
satisfy either
\begin{equation}
C_i = 0\;, \hs{1cm} (i=0,1,\dots,n)\;,
\end{equation}
which gives the Neumann boundary condition, or
\begin{equation}
C_i = \pm 1\;, \hs{1cm} (i=0,1,\dots,n)\;, \label{boundcond}
\end{equation}
which will be denoted as the (++...-- --...) boundary
conditions.

The original ideas needed to determine the full S-matrix in the
presence of integrable boundary conditions are due to
Cherednik \cite{chere84},
Ghoshal and Zamolodchikov\cite{ghosh94} and Fring and
K\"{o}berle\cite{fring94}.
As a consequence of the higher-spin symmetries of affine Toda theory
all amplitudes factorise into
products of the two-particle scattering amplitudes and the
single-particle reflection amplitudes.
These amplitudes in turn are strongly restricted by the
requirements of real analyticity, unitarity, crossing symmetry and
the bootstrap equations. We recall some details regarding this in
section \ref{sect:boot}.

If one knows the spectrum of
the integrable theory under investigation, these constraints are
usually strong enough to determine the amplitudes uniquely. This
explains the success in finding the S-matrices for affine Toda theories on the
whole line, because one can read off the particle spectrum directly
from the Lagrangian. In a theory on the half-line however a large
number of boundary bound states can exist in addition to the
particle states.
Unfortunately there is
no obvious way how to read off the spectrum of boundary states
directly from the Lagrangian.
This is the reason why many previous attempts to determine the
reflection amplitudes relying only on the above mentioned constraints
have failed\footnote{See however \cite{corri94,corri94b} in
which conjectures
were based on calculations of boundary solutions.
Some of these conjectures we will confirm to be correct.}.
We will see below that the correct reflection factors
have a rather intricate boundary state spectrum.

The new approach to finding the particle reflection amplitudes
which we will use in this paper has
been proposed in \cite{gande98b}. It is based on an earlier discovery
that the lowest breather states in imaginary coupling
Toda theory can be identified with the fundamental particles in
the real coupling theory in the sense that the particle
S-matrices are obtained from the breather \mbox{S-matrices} by
analytic continuation in the coupling constant \cite{gande95,gande96}.
Previously this phenomenon was only known to be exhibited by the
sine-Gordon model.
In \cite{gande98b} it was
proposed that the same should also hold for the reflection
amplitudes.
Analytic continuation of the breather reflection amplitudes should
give the particle reflection amplitudes.

The breather reflection amplitudes are determined by
bootstrap from the reflection amplitudes of the constituent solitons.
The soliton reflection amplitudes in turn can be obtained by
solving the reflection equation which is the analogue
of the Yang-Baxter equation for reflection amplitudes.
In \cite{gande98b} the reflection equation for $a_2^{(1)}$
Toda theory was solved and the particle
reflection amplitudes for
$a_2^{(1)}$ Toda theory with Neumann or $(+++)$ boundary condition
were calculated.
The results are briefly recalled
and then generalised to $a_n^{(1)}$ Toda theory with arbitrary $n$
in section \ref{sect:an}.
We obtain the reflection amplitudes for the uniform $(++\cdots ++)$
boundary condition in \eq{Ka} and find that they
confirm an early conjecture made in \cite{corri94}.
The reflection amplitudes for the Neumann
boundary condition, which we conjecture to be obtained by duality,
are given in \eq{KNeum}.

A prerequisite for progress towards deriving the reflection
amplitudes
for the other integrable boundary conditions was an understanding
of their classical vacuum solutions.
Bowcock \cite{bowco96b} determined these vacuum solutions for
$a_n^{(1)}$ Toda theories.
This study of classical solutions was simplified and generalised
in \cite{deliu98}. It was found that for about half of all integrable
boundary conditions in $a_n^{(1)}$ Toda theory real-valued vacuum solutions
can be obtained by analytically continuing a solution describing
a stationary soliton in front of the
boundary. We therefore call these boundary conditions `solitonic'.
Some details of this are given in section \ref{sect:vacuum}.
With this knowledge it is then easy to obtain the particle reflection
amplitudes for the vacuum states for these boundary conditions.
This is done in section \ref{sect:mixed}. We find that the
boundary conditions fall into $\lfloor(n+1)/2\rfloor$ equivalence
classes (where $\lfloor k\rfloor$ is the integer part of $k$).
Namely all boundary conditions whose vacuum solutions are
obtained from solitons in the same multiplet or from solitons in
the conjugate multiplet give rise to the same reflection
amplitudes. The result for the reflection amplitudes can be found
in eqs. \eqref{Kmixed}, \eqref{statsolfactor}.

In section \ref{sect:poles} we study the pole structure of these
reflection amplitudes.
This constitutes most of the work in this paper. We discover that
the vacuum reflection factors have simple poles which imply the
existence of excited boundary states. We can obtain the amplitudes
for reflection off these excited boundaries by using the
boundary bootstrap equations.
These in turn have new poles implying the existence of yet more
excited boundary states.

\begin{figure}
\begin{center}
\unitlength 0.8mm
\linethickness{0.4pt}
\begin{picture}(151.00,60.00)(0,50)
\put(80.00,100.00){\line(-1,-2){10.00}}
\put(70.00,80.00){\line(1,-2){10.00}}
\put(80.00,60.00){\line(1,2){10.00}}
\put(90.00,80.00){\line(-1,2){10.00}}
\put(100.00,65.00){\line(-2,3){10.00}}
\put(80.00,60.00){\line(4,1){20.00}}
\put(100.00,65.00){\line(4,3){20.00}}
\put(120.00,80.00){\line(1,1){20.00}}
\put(80.00,60.00){\line(-4,1){20.00}}
\put(60.00,65.00){\line(-4,3){20.00}}
\put(40.00,80.00){\line(-1,1){20.00}}
\put(60.00,65.00){\line(2,3){10.00}}
\put(80.00,100.00){\circle*{4.00}}
\put(90.00,80.00){\circle*{4.00}}
\put(70.00,80.00){\circle*{4.00}}
\put(20.00,100.00){\circle*{4.00}}
\put(40.00,80.00){\circle*{4.00}}
\put(60.00,65.00){\circle*{4.00}}
\put(80.00,60.00){\circle*{4.00}}
\put(100.00,65.00){\circle*{4.00}}
\put(120.00,80.00){\circle*{4.00}}
\put(140.00,100.00){\circle*{4.00}}
\put(11.00,61.00){\vector(0,1){49.00}}
\put(140.00,52.00){\line(0,-1){4.00}}
\put(130.00,52.00){\line(0,-1){4.00}}
\put(120.00,52.00){\line(0,-1){4.00}}
\put(110.00,52.00){\line(0,-1){4.00}}
\put(100.00,52.00){\line(0,-1){4.00}}
\put(90.00,52.00){\line(0,-1){4.00}}
\put(80.00,52.00){\line(0,-1){4.00}}
\put(70.00,52.00){\line(0,-1){4.00}}
\put(60.00,52.00){\line(0,-1){4.00}}
\put(50.00,52.00){\line(0,-1){4.00}}
\put(40.00,52.00){\line(0,-1){4.00}}
\put(30.00,52.00){\line(0,-1){4.00}}
\put(20.00,52.00){\line(0,-1){4.00}}
\put(80.00,104.00){\makebox(0,0)[cb]{\shs{\scs{$b_{1,1}$}}}}
\put(140.00,104.00){\makebox(0,0)[cb]{\shs{\scs{$b_{-3,3}$}}}}
\put(20.00,104.00){\makebox(0,0)[cb]{\shs{\scs{$b_{3,-3}$}}}}
\put(36.00,77.00){\makebox(0,0)[rt]{\shs{\scs{$b_{-2,2}$}}}}
\put(124.00,77.00){\makebox(0,0)[lt]{\shs{\scs{$b_{2,-2}$}}}}
\put(65.00,80.00){\makebox(0,0)[rc]{\shs{\scs{$b_{0,1}$}}}}
\put(95.00,80.00){\makebox(0,0)[lc]{\shs{\scs{$b_{1,0}$}}}}
\put(60.00,60.00){\makebox(0,0)[ct]{\shs{\scs{$b_{-1,1}$}}}}
\put(100.00,60.00){\makebox(0,0)[ct]{\shs{\scs{$b_{1,-1}$}}}}
\put(80.00,55.00){\makebox(0,0)[ct]{\shs{\scs{$b_{0,0}$}}}}
\put(88.00,91.00){\makebox(0,0)[lc]{\shs{\scs{$2$}}}}
\put(77.00,71.00){\makebox(0,0)[lc]{\shs{\scs{$2$}}}}
\put(98.00,72.00){\makebox(0,0)[lc]{\shs{\scs{$2$}}}}
\put(72.00,91.00){\makebox(0,0)[rc]{\shs{\scs{$1$}}}}
\put(62.00,72.00){\makebox(0,0)[rc]{\shs{\scs{$1$}}}}
\put(83.00,71.00){\makebox(0,0)[rc]{\shs{\scs{$1$}}}}
\put(68.00,60.00){\makebox(0,0)[ct]{\shs{\scs{$1$}}}}
\put(46.00,70.00){\makebox(0,0)[ct]{\shs{\scs{$1$}}}}
\put(27.00,89.00){\makebox(0,0)[rt]{\shs{\scs{$1$}}}}
\put(90.00,60.00){\makebox(0,0)[ct]{\shs{\scs{$2$}}}}
\put(112.00,70.00){\makebox(0,0)[ct]{\shs{\scs{$2$}}}}
\put(132.00,88.00){\makebox(0,0)[lt]{\shs{\scs{$2$}}}}
\put(6.00,110.00){\makebox(0,0)[rc]{$E$}}
\put(15.00,50.00){\vector(1,0){136.00}}
\put(146.00,46.00){\makebox(0,0)[ct]{$n_1-n_2$}}
\end{picture}
\end{center}
\caption{\it {
The spectrum of boundary states for $a_2^{(1)}$ Toda
theory with a solitonic boundary condition at coupling constant
$B\approx\frac{1}{2}$. The vertical axis is the boundary state
energy.}
\label{spectrum1}}
\end{figure}
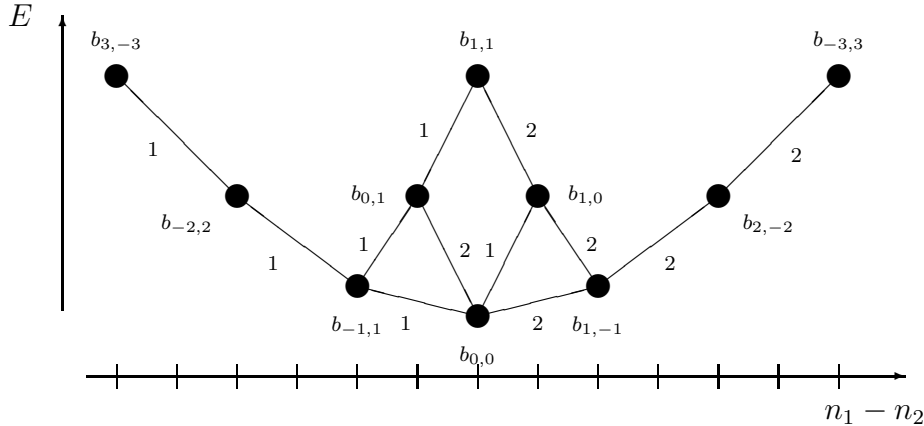

\begin{figure}
\begin{center}
\epsfbox{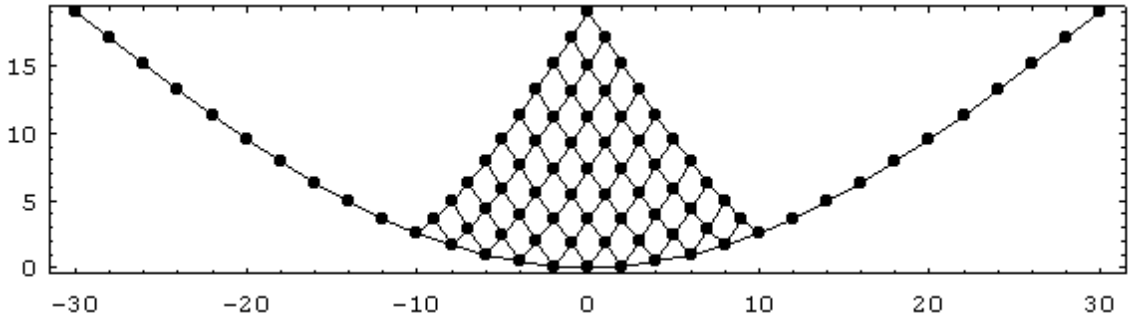}
\end{center}
\caption{\it
As the coupling constant becomes smaller, the number of boundary
states grows. This plot shows the spectrum as in figure
\ref{spectrum1} but at coupling constant
$B\approx\frac{1}{10}$.
\label{spectrum2}}
\end{figure}

We have performed this analysis in detail
for the cases of the $a_2^{(1)}$
and $a_4^{(1)}$ Toda theories. In each of these cases we
end up with an array of
boundary states. The number of excited boundary states grows as
the coupling constant becomes smaller. As an example we have
plotted in figures \ref{spectrum1} and \ref{spectrum2} the
spectrum of boundary states for the solitonic boundary conditions
of $a_2^{(1)}$ affine Toda theory at two different values of the
coupling constant
$B=\frac{\beta^2}{2\pi}/(1+\frac{\beta^2}{4\pi})$. In
figure~\ref{spectrum1} we have also indicated which particle
has to bind to the
boundary to go from one state to the other by labels on
the lines connecting the states. A more illuminating animated plot
of the spectrum can be found on this paper's web site at
{\tt http//www.mth.kcl.ac.uk/\~{}delius/pub/reflection.html}.

The amplitudes
for the reflection of the particles off these boundary states are
given explicitly in \eq{Khnm} for $a_2^{(1)}$, in eqs. \eqref{k11}
and \eqref{k21} for the first class of boundary conditions in
$a_4^{(1)}$ and eqs. \eqref{k12} and \eqref{k22} for the second
class. We have also begun the analysis for $a_3^{(1)}$ in section
\ref{a3}.

In these reflection amplitudes there are also a large number of
physical
strip poles which do not correspond to new bound states but have
an explanation in terms of a boundary generalisation \cite{dorey98}
of the Coleman-Thun mechanism \cite{cole}. The way in which this works
conveys the strong impression that there is some magic at
work which awaits a deeper explanation.

\section{Exact S-matrices and reflection amplitudes\label{sect:boot}}

This section gives a brief review of the properties of
particle S-matrices and reflection amplitudes and the consistency
equations which they are subject to. For an excellent presentation of
this material see \cite{corri96}.

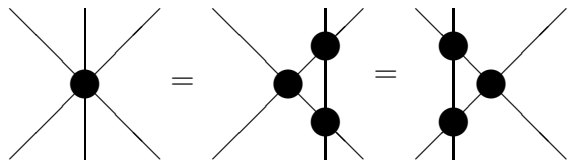
\begin{figure}[b]
\unitlength 1mm
\linethickness{0.4pt}
\begin{picture}(84.00,30.00)(-30,8)
\put(10.00,10.00){\line(1,1){20.00}}
\put(10.00,30.00){\line(1,-1){20.00}}
\put(69.00,10.00){\line(0,1){20.00}}
\put(37.00,10.00){\line(1,1){20.00}}
\put(64.00,10.00){\line(1,1){20.00}}
\put(37.00,30.00){\line(1,-1){20.00}}
\put(64.00,30.00){\line(1,-1){20.00}}
\put(52.00,10.00){\line(0,1){20.00}}
\put(20.00,10.00){\line(0,1){20.00}}
\put(33.00,20.00){\makebox(0,0)[cc]{$=$}}
\put(60.00,21.00){\makebox(0,0)[cc]{$=$}}
\put(20.00,20.00){\circle*{4.00}}
\put(47.00,20.00){\circle*{4.00}}
\put(74.00,20.00){\circle*{4.00}}
\put(52.00,25.00){\circle*{4.00}}
\put(52.00,15.00){\circle*{4.00}}
\put(69.00,25.00){\circle*{4.00}}
\put(69.00,15.00){\circle*{4.00}}
\end{picture}
\caption{\it Higher-spin symmetries allow one to shift the different
particle trajectories independently, thereby separating
multi-particle scattering processes into products of two-particle
scatterings.
\label{factorization}}
\end{figure}

The reason why the full S-matrices of affine Toda theories can be
determined exactly is that these theories possess
higher-spin symmetries.
A higher-spin symmetry implies that the sum of some
power of the momenta of the particles is conserved during scattering.
This allows only processes in which the incoming particles have
the same momenta as the outgoing ones. Furthermore
a higher-spin symmetry acts on the particles by shifting them
by amounts proportional to powers of their momenta.
This allows one to separate
the interaction regions as illustrated in figure
\ref{factorization}. This implies that any S-matrix factorizes
into a product of 2-particle S-matrices.

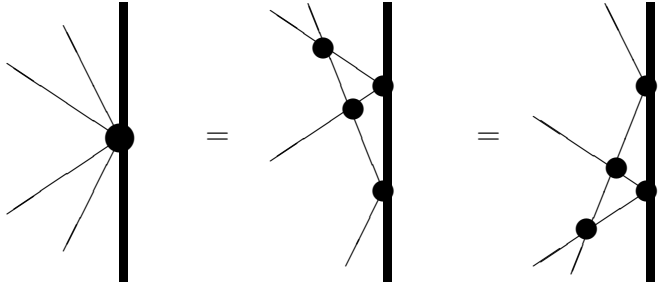
\begin{figure}
\unitlength 1.00mm
\linethickness{0.4pt}
\begin{picture}(121.42,36.00)(0,18)
\put(120.00,28.00){\line(-3,-2){15.00}}
\put(120.00,28.00){\line(-3,2){15.00}}
\put(120.00,42.00){\line(-2,-5){10.00}}
\put(85.00,28.00){\line(-2,5){10.00}}
\put(85.00,42.00){\line(-3,-2){15.00}}
\put(85.00,42.00){\line(-3,2){15.00}}
\put(50.00,35.00){\line(-3,-2){15.00}}
\put(50.00,35.00){\line(-3,2){15.00}}
\put(50.00,35.00){\circle*{4.00}}
\put(77.00,47.00){\circle*{2.83}}
\put(85.00,42.00){\circle*{2.83}}
\put(81.00,39.00){\circle*{2.83}}
\put(85.00,28.00){\circle*{2.83}}
\put(120.00,42.00){\circle*{2.83}}
\put(116.00,31.00){\circle*{2.83}}
\put(120.00,28.00){\circle*{2.83}}
\put(112.00,23.00){\circle*{2.83}}
\put(63.00,35.00){\makebox(0,0)[cc]{$=$}}
\put(99.00,35.00){\makebox(0,0)[cc]{$=$}}
\put(85.00,28.00){\line(-1,-2){5.00}}
\put(120.00,42.00){\line(-1,2){5.50}}
\put(50.00,35.00){\line(-1,-2){7.50}}
\put(50.00,35.00){\line(-1,2){7.50}}
\put(85.00,16.00){\rule{3pt}{37mm}}
\put(120.00,16.00){\rule{3pt}{37mm}}
\put(50.00,16.00){\rule{3pt}{37mm}}
\end{picture}
\caption{\it In the presence of an integrable boundary any
multi-particle process factorises into a product of two-particle
scatterings and single-particle reflections.
\label{bfact}}
\end{figure}

If the particles live on a half-line, then in addition to
scattering with each other they are also going to reflect off the
boundary. Ideally one would hope that a multi-particle process
will now factorise into a succession of two-particle scatterings
and single-particle reflections off the boundary, as depicted in
figure~\ref{bfact}.
Of course all momentum-dependent translation
symmetries in the space direction will be broken by the boundary.
However one can choose the boundary condition in such a way that
it preserves those parts of the higher-spin-symmetries which
effect momentum-dependent translations in the time direction and
these suffice to separate the interaction regions. Thus in order to
determine the full S-matrix on the half-line we only need
the single-particle reflection matrices and the two-particle
S-matrices. The later will not be changed by the presence of the
boundary because
the particle scatterings could be shifted arbitrarily far away
from the boundary.

In affine Toda theory all particles are distinguished by their
masses and higher-spin charges and therefore the S-matrix has to
be diagonal.
\begin{equation}
\unitlength 1mm
\linethickness{0.4pt}
\begin{picture}(47.00,33.00)(20,0)
\put(25.00,5.00){\line(1,1){20.00}}
\put(25.00,25.00){\line(1,-1){20.00}}
\put(35.00,15.00){\circle*{2.00}}
\put(25.00,2.00){\makebox(0,0)[ct]{\scs{$a$}}}
\put(45.00,2.00){\makebox(0,0)[ct]{\scs{$b$}}}
\put(25.00,27.00){\makebox(0,0)[cb]{\scs{$c$}}}
\put(45.00,27.00){\makebox(0,0)[cb]{\scs{$d$}}}
\put(35.00,8.00){\makebox(0,0)[ct]{\scs{$i(\th_1-\th_2)$}}}
\put(22.00,15.00){\makebox(0,0)[rc]{$S_{ab}^{cd}(\theta_1,\theta_2)\ =$}}
\put(47.00,15.00){\makebox(0,0)[lc]
{$=\ \delta_{bc}\,\delta_{ad}\,S_{ab}(\theta_1-\theta_2)$.}}
\put(35.00,12.50){\oval(4.00,3.00)[b]}
\end{picture}
\vspace{5mm}
\end{equation}
We have introduced the rapidity variable $\th$ which is
related to the momentum components by
\begin{equation}
E=m\cosh\th, ~~~~p=m\sinh\th,
\end{equation}
where $m$ is the mass of the particle.
We draw the worldline of a particle with rapidity $\theta$ at an
angle of $u=-i\theta$ to the vertical.

Depending on which of the higher-spin symmetries
remain unbroken by the boundary condition a particle has to be
reflected either into itself or into its antiparticle.
We will find that for the boundary conditions which we study
the former is the case and so the reflection matrices are diagonal
\begin{equation}
\unitlength 1.00mm
\linethickness{0.4pt}
\begin{picture}(45.00,25.00)(20,18)
\put(40.00,20.00){\rule{3pt}{20.00mm}}
\put(30.00,20.00){\line(1,1){10.00}}
\put(40.00,30.00){\line(-1,1){10.00}}
\put(40.00,30.00){\circle*{2.00}}
\put(38.50,26.50){\oval(3.00,3.00)[b]}
\put(37.00,21.00){\makebox(0,0)[cc]{$i\th$}}
\put(27.00,30.00){\makebox(0,0)[rc]{$K_a^b(\th)\ =$}}
\put(45.00,30.00){\makebox(0,0)[lc]{$=\ \delta_{ab}\,K_a(\th)$}}
\put(28.00,19.00){\makebox(0,0)[lt]{\scs{$a$}}}
\put(28.00,41.00){\makebox(0,0)[lb]{\scs{$a$}}}
\end{picture}
\end{equation}

The scattering and the reflection amplitudes have to be unitary,
real analytic and $2\pi i$-periodic. This implies that they
can be written as a
product of  fundamental building blocks (we use the notation of
\cite{brade90} in which $h=n+1$ is the Coxeter number of $a_n^{(1)}$)
\begin{equation}\label{bracketre}
\bl x\br\equiv\frac{\sin\left(\frac{\th}{2i}+\frac{\pi}{2h}x\right)}
{\sin\left(\frac{\th}{2i}-\frac{\pi}{2h}a\right)},
\end{equation}
\begin{equation}
S_{ab}(\th)=\prod_{x\in A_{ab}}\bl x\br\,,~~~~~~
K_a(\th)=\prod_{x\in B_a}\bl x\br\;.
\end{equation}
Each block $(x)$ has a pole at $\th=x\,i\pi/h$ and a zero at
$\th=-x\,i\pi/h$.
Thus the problem of calculating the full exact S-matrix has been reduced
to the task of finding the sets $A_{ab}$ and $B_a$ of numbers specifying the
locations of the poles and zeros in the two-particle scattering amplitudes
$S_{ab}$ and the reflection amplitudes $K_a$ respectively.

As an illustration we give the scattering amplitudes for $a_2^{(1)}$
affine Toda theory which we will meet again in section \ref{subsect:mixed2}.
This theory has two conjugate particles of equal mass with
scattering amplitudes
\begin{gather}
S_{11}=S_{22}=\bl -2+B\br\,\bl -B\br\,\bl 2\br\,,\label{S11}\\
S_{12}=S_{21}=-\,\bl -3+B\br\,\bl -1-B\br\,\bl 1\br\,.\label{S12}
\end{gather}
Here $B$ is the usual coupling
constant dependent function
\begin{equation}\label{B}
B(\beta) =\frac 1{2\pi} \frac{\beta^2}{1+\frac{\beta^2}{4\pi}}\;,
\end{equation}
The reflection amplitudes depend on which integrable boundary
condition we choose. For example for the solitonic boundary
condition dealt with in section \ref{subsect:mixed2} we have
\begin{equation}\label{K1}
K_1=K_2=\bl -2\br\,\bl -\frac B2\br\,\bl 2+\frac B2 \br\,
\bl \frac 12 - \frac B2\br\, \bl
\frac 32 - \frac B2 \br\, \bl \frac 32 + \frac B2 \br\, \bl \frac 52 +
\frac B2 \br\;.
\end{equation}
Crossing symmetry relates different amplitudes to each other,
\begin{equation}
S_{ab}(\th)=S_{b\bar{a}}(i\pi-\th),~~~~~
K_a(\th)K_{\bar a}(\th+i\pi)=S_{aa}(2\th),
\end{equation}
where $\bar{a}$ denotes the antiparticle to $a$ and is given for
$a_n^{(1)}$ Toda theory by $\bar{a}=h-a$.
The second of these relations was first derived by Ghoshal and
Zamolodchikov \cite{ghosh94}. It can be checked that the example given
above satisfies these crossing relations by using the following properties
of the blocks:
\begin{equation}
\bl a\br_{\th+i\pi}=-\bl a+h\br,~~~~
\bl a\br_{2\th}=-\bl\frac{a}{2}\br\,\bl\frac a2 +h\br,~~~~
\bl a\br_{-\th}=-\bl a\br^{-1}.
\end{equation}

\begin{figure}
\unitlength 0.50mm
\linethickness{0.4pt}
\begin{picture}(154.00,61.00)(-35,10)
\put(30.00,30.00){\line(2,1){20.00}}
\put(50.00,40.00){\line(2,-1){20.00}}
\put(50.00,40.00){\line(0,1){20.00}}
\put(30.00,70.00){\line(2,-1){20.00}}
\put(50.00,60.00){\line(2,1){20.00}}
\put(35.00,29.00){\makebox(0,0)[lt]{\scs{$1$}}}
\put(65.00,29.00){\makebox(0,0)[rt]{\scs{$1$}}}
\put(65.00,71.00){\makebox(0,0)[rb]{\scs{$1$}}}
\put(35.00,71.00){\makebox(0,0)[lb]{\scs{$1$}}}
\put(52.00,50.00){\makebox(0,0)[lc]{\scs{$2$}}}
\put(50.00,34.00){\makebox(0,0)[cc]{\scs{$\frac{2 i\pi}{3}$}}}
\put(150.00,30.00){\rule{3pt}{20mm}}
\put(150.00,42.00){\line(-5,-3){20.00}}
\put(150.00,58.00){\line(-5,3){20.00}}
\put(154.00,35.00){\makebox(0,0)[lc]{\scs{vacuum}}}
\put(154.00,50.00){\makebox(0,0)[lc]{\scs{excited boundary}}}
\put(154.00,65.00){\makebox(0,0)[lc]{\scs{vacuum}}}
\put(135.00,29.00){\makebox(0,0)[lt]{\scs{$1$}}}
\put(135.00,71.00){\makebox(0,0)[lb]{\scs{$1$}}}
\put(50.00,20.00){\makebox(0,0)[cc]{a)}}
\put(150.00,20.00){\makebox(0,0)[cc]{b)}}
\end{picture}
\caption{\it a) Simple poles in a scattering amplitude may be due to
the propagation of a bound state. b) Simple poles in a reflection
amplitude may be due to the formation of an excited boundary
state.
\label{fig:pole}}
\end{figure}
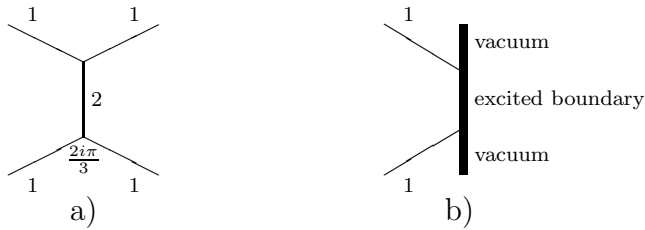

By going from the Mandelstamm variable $s$ to the relative rapidity variable
$\th$ the physical sheet has been mapped to the physical strip
$0<Im(\th)<\pi$ (see for example \cite{Zam} for details).
Every pole on the physical strip has to have a physical
explanation. In the example above the amplitude $S_{11}$ has a
pole from the block $(2)$ at $\theta=2 i\pi/3$ which
corresponds to the propagation of a particle of type $2$ in the forward
channel, as depicted in figure~\ref{fig:pole} a).

A further consequence of the higher-spin symmetries and the
corresponding momentum-dependent translation of particle lines is
the bootstrap principle. It
expresses the scattering amplitudes for bound states in terms of
those of the constituent particles. For example the bound state
pole discussed above leads to the following bootstrap equation
which gives $S_{12}$ in terms of $S_{11}$,
\begin{equation}
\unitlength 0.6mm
\linethickness{0.4pt}
\begin{picture}(185.00,45.00)(0,10)
\put(20.00,20.00){\line(2,1){30.00}}
\put(50.00,35.00){\line(2,-1){30.00}}
\put(50.00,35.00){\line(0,1){16.00}}
\put(123.00,20.00){\line(2,1){30.00}}
\put(153.00,35.00){\line(2,-1){30.00}}
\put(153.00,35.00){\line(0,1){15.00}}
\put(117.00,24.00){\line(6,1){68.00}}
\put(14.00,38.00){\line(6,1){68.00}}
\put(160.00,31.00){\circle*{4.00}}
\put(138.00,27.00){\circle*{4.00}}
\put(50.00,44.00){\circle*{4.00}}
\put(20.00,17.00){\makebox(0,0)[ct]{\scs{$1$}}}
\put(80.00,17.00){\makebox(0,0)[ct]{\scs{$1$}}}
\put(123.00,16.00){\makebox(0,0)[ct]{\scs{$1$}}}
\put(183.00,16.00){\makebox(0,0)[ct]{\scs{$1$}}}
\put(50.00,53.00){\makebox(0,0)[cb]{\scs{$2$}}}
\put(153.00,53.00){\makebox(0,0)[cb]{\scs{$2$}}}
\put(115.00,24.00){\makebox(0,0)[rc]{\scs{$2$}}}
\put(12.00,38.00){\makebox(0,0)[rc]{\scs{$2$}}}
\put(56.00,41.00){\makebox(0,0)[lt]{\fns{$S_{12}$}}}
\put(138.00,22.00){\makebox(0,0)[ct]{\fns{$S_{11}$}}}
\put(160.00,26.00){\makebox(0,0)[ct]{\fns{$S_{11}$}}}
\put(96.00,34.00){\makebox(0,0)[cc]{$=$}}
\end{picture}
\end{equation}
This provides an independent consistency check on the amplitude $S_{12}$
given in \eqref{S12} which was already determined by crossing
symmetry. In practice there is a large number of such bootstrap
consistency equations and these, together with the requirement
that any pole on the physical strip needs to have a physical
explanation, are strong enough to determine the S-matrices of all
affine Toda theories uniquely.

Similar relations exist for the reflection amplitudes.
Firstly there is a bootstrap equation relating the reflection
amplitudes of the different particles to each other, depicted in
figure~\ref{figbootstrap}. Secondly
every pole on the physical strip requires a physical
explanation. For reflection amplitudes the physical strip in the
rapidity plane is bounded by $Im(\th)\leq\pi/2$.
Our example in \eqref{K1} has two simple poles on the physical
strip, namely at $\th=i\frac{\pi}{3}\left(\frac 12-\frac B2\right)$
and at $\th=i\frac{\pi}{3}\left(\frac 32-\frac B2\right)$. These
correspond to the propagation of excited boundary states as
illustrated in figure~\ref{fig:pole} b).
The reflection amplitudes for reflection of these excited
boundaries is determined by the following bootstrap
equation,
\begin{equation}\label{bb}
\unitlength 0.5mm
\linethickness{0.4pt}
\begin{picture}(122.00,45.00)(10,8)
\put(50.00,10.00){\rule{3pt}{20.00mm}}
\put(50.00,30.00){\line(-4,-5){16.00}}
\put(120.00,25.00){\line(-3,-1){30.00}}
\put(120.00,25.00){\line(-3,1){30.00}}
\put(120.00,10.00){\rule{3pt}{20.00mm}}
\put(120.00,40.00){\line(-3,-4){22.50}}
\put(50.00,40.00){\line(-3,-1){30.00}}
\put(50.00,40.00){\line(-3,1){30.00}}
\put(50.00,40.00){\circle*{4.00}}
\put(111.00,28.00){\circle*{4.00}}
\put(120.00,25.00){\circle*{4.00}}
\put(105.00,20.00){\circle*{4.00}}
\put(70.00,30.00){\makebox(0,0)[cc]{$=$}}
\end{picture}
\end{equation}
The bootstrap equations would probably again be strong enough to
determine the reflection amplitudes uniquely if one knew the
spectrum of excited boundary states a priori. However this is not
easy to derive directly and therefore we took a different route,
to be described below.

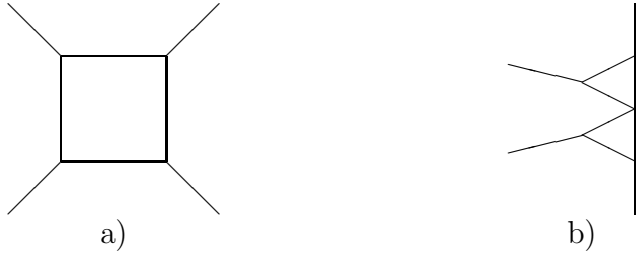
\begin{figure}
\begin{center}
\unitlength 0.7mm
\linethickness{0.4pt}
\begin{picture}(149.00,50.00)
\put(40.00,40.00){\line(1,0){20.00}}
\put(60.00,40.00){\line(0,-1){20.00}}
\put(60.00,20.00){\line(-1,0){20.00}}
\put(40.00,20.00){\line(0,1){20.00}}
\put(40.00,40.00){\line(-1,1){10.00}}
\put(60.00,40.00){\line(1,1){10.00}}
\put(60.00,20.00){\line(1,-1){10.00}}
\put(40.00,20.00){\line(-1,-1){10.00}}
\put(149.00,50.00){\line(0,-1){40.00}}
\put(149.00,40.00){\line(-2,-1){10.00}}
\put(139.00,35.00){\line(2,-1){10.00}}
\put(149.00,30.00){\line(-2,-1){10.00}}
\put(139.00,25.00){\line(2,-1){10.00}}
\put(139.00,35.00){\line(-4,1){14.00}}
\put(139.00,25.00){\line(-4,-1){14.00}}
\put(50.00,6.00){\makebox(0,0)[cc]{a)}}
\put(139.00,6.00){\makebox(0,0)[cc]{b)}}
\end{picture}
\caption{\it Diagrams with internal lines on shell lead to additional
poles on the physical strip.\label{ct}}
\end{center}
\end{figure}

Not only tree diagrams as those in figure~\ref{fig:pole} give rise to
poles on the physical strip. Rather any Feynman diagram in which
some internal lines are on-shell can produce poles. For example
the box diagram in figure~\ref{ct}a) will lead to a second order
pole if all the internal lines can be simultaneously on shell
(which does not happen in $a_2^{(1)}$ but in higher
Toda theories). The order of the pole is determined by the number
of lines on shell minus twice the number of loop integrations. The
boundary diagram in figure~\ref{ct}(b) has six internal lines and
two loop momenta and will thus generically lead to a second order
pole unless the reflection amplitude in the middle of the diagram
does itself have poles or zeros at this point. This boundary
generalisation of the Coleman-Thun mechanism \cite{cole} was discovered by
Dorey et.al. in \cite{dorey98}.

Thus to check a conjectured amplitude one should draw all possible
diagrams which can have internal lines on shell, determine what
kind of poles they lead to, and then compare with the pole
structure of the conjectured amplitude. These checks have been
carefully performed for the scattering amplitudes of affine Toda
theory on the whole line. We will do the same for our reflection
amplitudes for $a_2^{(1)}$ and $a_4^{(1)}$ Toda theory in section
\ref{sect:poles} and will find that it works quite miraculously.
A similar analysis has been performed for the
reflection amplitudes in the boundary scaling Lee-Yang model
\cite{dorey98}.
\section{Reflection amplitudes for Neumann and $(++\cdots ++)$
boundary conditions\label{sect:an}}
\subsection{Reflection in $\ato$ ATFT revisited}

In a recent paper \cite{gande98b} one of us constructed reflection
amplitudes for $\ato$ ATFT. We briefly recall the main results of this
work here.

The construction was based on new solutions to the $\ato$
boundary Yang-Baxter equation (BYBE).
Some of these new solutions, the so-called $K$-matrices, were conjectured to
describe the reflection of the fundamental solitons in imaginary
coupling $\ato$ ATFT.
By comparison with semi-classical calculations \cite{deliu98} it was
possible to identify the two reflection matrices which correspond
to the Neumann and
the uniform (+++) boundary conditions.
Using the boundary bootstrap equations,
the reflection amplitudes for the breathers in
the theory were constructed.
Due to the identification of the lowest breathers in the imaginary
coupling theory with the fundamental particles in the real coupling
theory, it was then possible to find the reflection amplitudes for the
particles in the real coupling $\ato$ ATFT via a simple analytic
continuation.

This led to the following particle reflection amplitudes,
corresponding to the Neumann and the uniform (+++) boundary conditions
respectively,
\bea
K^{N}_1(\th) = K^{N}_2(\th)&=&\bl -2\br\,
\bl -\frac B2\br\, \bl 2+\frac B2\br \;,  \\ \nn \\
K_1(\th) = K_2(\th)&=&\bl
-2\br\, \bl \frac B2-1\br\, \bl 3-\frac B2\br \;, \label{realref1}
\eea
in which the brackets are defined as in (\ref{bracketre}) with $h=3$.
These amplitudes are in agreement with classical and semi-classical
calculations \cite{corri94b,kim95b,perki98}.

Given that the $S$-matrices in the bulk theory have long been known to
be self-dual under a weak--strong coupling duality, it was widely
expected that a duality like that can also be found in ATFT on a
half-line. However, the above reflection amplitudes are
not self-dual, but instead are dual to each other, i.e.
\begin{equation}
K^{N}_1(\th) \longleftrightarrow
K_1(\th) \;,
\hs{2cm} (\mbox{as}\;\;\; \beta
\longleftrightarrow  \frac{4\pi}{\beta})\;.
\end{equation}
This generalises the analogous
result in the sinh-Gordon theory in \cite{corri97}.

In the following we demonstrate how these results can be
generalised to the case of $\ao$ ATFTs.
Just as in the $\ato$ case, our construction of particle reflection
amplitudes begins with proposing a reflection matrix for the fundamental
solitons in imaginary coupling ATFT with uniform $(++\cdots ++)$
boundary condition.

\subsection{Notation}
We use the notation of \cite{gande96}.
Whenever we deal with solitons or breathers, we use the
following parametrisation of the rapidity:
\[
\mu = -i \frac{h\la}{2\pi}\th\;,
\]
and the bracket notation
\begin{equation}
\bl a \bri
\equiv \frac{\lb a \rb} {\lb -a \rb}\;,
\;\;\;\;\;\;\;\;\;\;
\lb a \rb \equiv \sin\left(\frac{\pi}{h\la}(\mu +a)\right)
\;, \label{bracketim}
\end{equation}
in which $h=n+1$ is the Coxeter number of the affine Lie algebra
$\ao$, and $\la$ is related to the imaginary Toda
coupling constant $\beta$ in the following way:
\be\label{lambda}
\la = -\frac{4\pi}{\beta^2} - 1\;.
\end{equation}
This is related to the function $B(\beta)$ in \eq{B} which is used
in real coupling Toda theory by $\lambda=-2/B$.
We introduced the subscript $I$ in order to distinguish this
block  notation from the block notation of \eq{bracketre}
which we use when dealing with the fundamental particles.

\subsection{The diagonal $\ao$ soliton reflection matrices}

Although new non-diagonal solutions to the $\ao$ BYBE have been found
recently \cite{gande99}, it will turn out that here we only need the
diagonal solutions.
In these \mbox{$K$-matrices} all non-diagonal entries are zero, which
significantly simplifies all other equations, like the bootstrap,
crossing and unitarity conditions.

We know from the classical calculations in \cite{deliu98} that
solitons are always reflected into antisolitons by the boundaries
which we are considering\footnote{Also a soliton-preserving boundary
conditions exists \cite{deliu98b} but we do not consider it
here.}.
We are therefore only interested in those solutions of the BYBE in which
the solitons change into antisolitons, which means the $K$-matrices
are maps:
\be
K_{A^{(a)}}(\mu) : V_a \lra V_{n+1-a}\;, \;\;\;\;\;\;\;\; (a = 1,2,...,n)\;.
\end{equation}
It is fairly straightforward to show that there is a solution to the
$\ao$ BYBE of the form
\be
K_{A^{(1)}}(\mu) = \left( \begin{array}{cccccc}
1 & 0 & \dots & & & \\
0 & 1 &  & & & \\
\vdots & & \ddots & & & \\
& & & & & \\
& & & & & 1 \end{array} \right)\, {\cal A}_1(\mu)\;,
\end{equation}
which is a $n\times n$ square matrix, which maps the solitons in the
first multiplet into their charge conjugate partners in the $n$th
multiplet. ${\cal A}_1(\mu)$ is an overall scalar factor, which is
determined by the requirements of boundary unitarity and boundary
crossing.

Similarly as in the $\ato$ case from \cite{gande98b} the two
conditions of boundary unitarity and boundary crossing lead to the
following equations for ${\cal A}_1(\mu)$:
\bea\label{aeq}
{\cal A}_1(\mu)\, \ol{\cal A}_1(-\mu) &=& 1\;, \nn \\ \nn \\
\frac{{\cal A}_1(-\mu+\frac{n+1}4 \la)}{{\cal A}_1(\mu+\frac{n+1}4
\la)} &=& \frac{\ol{\cal A}_1(-\mu+\frac{n+1}4 \la)}{\ol{\cal
A}_1(\mu+\frac{n+1}4 \la)} = F_{1,1}(2\mu)\;,
\end{eqnarray}
in which $\ol{\cal A}_1 = {\cal A}_n$ is the overall scalar
factor of the charge conjugate reflection matrix, and
$F_{1,1}(\mu)$ is the scalar factor of the $\ao$ $S$-matrix
which was given in \cite{gande96}.
We can solve these equations in the same way as it was done in
\cite{gande98b}, and we obtain the minimal solution
\bea\label{aa}
{\cal A}_1(\mu) = \ol{\cal A}_1(\mu) = \bl - \frac{n+1}4 \la \bri\,
\prod_{k=1}^{\infty}
\frac{G_{2k}(2\mu, - \frac 32 (n+1)\la +1 )\, G_{2k}(2\mu, -\frac 12
(n+1) \la)} {G_{2k}(2\mu, - \frac 32 n\la -\frac{\la}2 +1
)\, G_{2k}(2\mu, -\frac 12 n\la -\frac 32 \la)}\;,
\end{eqnarray}
in which
\be
G_j(\mu,a) \equiv \frac{\Ga(\mu+jh\la+a)} {\Ga(-\mu+jh\la+a)}\;.
\end{equation}
There is the freedom to add certain CDD factors to this solution
as described in subsection \ref{scdd}.
We have checked that these soliton reflection amplitudes agree
semiclassically with the soliton time-delays calculated in
\cite{deliu98} for the uniform $(++\cdots++)$ boundary condition.

In principle it would now be possible to obtain the other soliton $K$-matrices
$K_{A^{(a)}}(\mu)$ for $(a=1,\dots,n)$ by using the bootstrap
equations. However, since we are mainly interested in the reflection
amplitudes for the breathers and subsequently for the particles in the
real coupling theory, we can first construct the reflection amplitudes
for the particle of type $1$ and then use the bootstrap equations
directly in the real coupling theory.

\subsection{The breather and particle reflection amplitudes}

The scattering amplitude for two solitons of type $1$ and type $n$ ($=\ol 1$)
has a pole at $\mu=\frac{n+1}2\lambda-1$ corresponding to the formation of the
lowest breather state $B_1^{(1)}$. This  leads to the
breather bootstrap equations which were described in great detail in
the appendix of \cite{gande98b}\footnote{For the sine-Gordon model the breather
reflection amplitudes were obtained by Ghoshal in \cite{ghosh94b}.}.
Without going into further detail
we can write down the result for the breather reflection
amplitude,
\bea\label{kb1}
K_{B_1^{(1)}}(\mu) &=& {\cal A}\left(\mu-\frac{n+1}4 \la+\frac 12\right)\,
F_{1,1}(2\mu)\, {\cal A}\left(\mu+\frac{n+1}4 \la-\frac 12\right) \nn \\
&=& \bl -\frac n2 \la \bri\, \bl -\frac 12(n+1) \la+\frac 12 \bri\,
\bl -\frac{\la}2-\frac 12 \bri\;. \label{KB1}
\end{eqnarray}
The fact that solitons reflect into antisolitons in the charge conjugate
multiplet ensures that the breather reflection is purely diagonal.

\begin{figure}[ht]
%
%
\begin{center}
\begin{picture}(330,190)(0,0)
\put(120,10){\rule{3pt}{140pt}}
\put(120,80){\line(-1,-1){30}}
\put(90,50){\line(-2,-1){60}}
\put(90,50){\line(-1,-2){15}}
\put(120,80){\line(-1,1){30}}
\put(90,110){\line(-2,1){60}}
\put(90,110){\line(-1,2){15}}
\put(160,78){\line(1,0){13}}
\put(160,82){\line(1,0){13}}
\put(300,10){\rule{3pt}{140pt}}
\put(300,80){\line(-4,-3){80}}
\put(300,80){\line(-4,3){24}}
\put(300,50){\line(-1,-2){15}}
\put(300,50){\line(-1,2){24}}
\put(276,98){\line(-1,1){26}}
\put(250,124){\line(-2,1){35}}
\put(250,124){\line(-1,2){10}}
\put(24,11){\shs{\fns{$a$}}}
\put(70,11){\shs{\fns{$b$}}}
\put(213,11){\shs{\fns{$a$}}}
\put(280,11){\shs{\fns{$b$}}}
\put(90,68){\shs{\scs{$a+b$}}}
\put(265,111){\shs{\scs{$a+b$}}}
\put(277,81){\shs{\scs{$b$}}}
\put(289,89){\shs{\scs{$a$}}}
\put(289,55){\shs{\scs{$b$}}}
\put(24,143){\shs{\fns{$b$}}}
\put(70,143){\shs{\fns{$a$}}}
\put(208,143){\shs{\fns{$b$}}}
\put(238,146){\shs{\fns{$a$}}}
\end{picture}
\end{center}
\caption{\it The boundary bootstrap relation}\label{figbootstrap}
\end{figure}
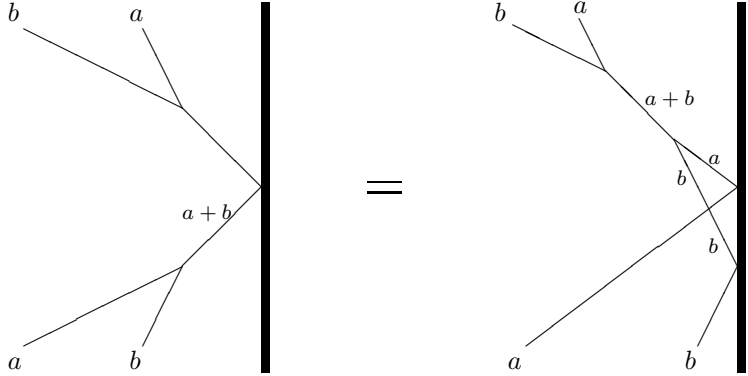

Since the breather $B_1^{(1)}(\mu)$ corresponds to the first particle
in real coupling $\ao$ ATFT, we obtain the reflection amplitude for
this particle through analytic continuation $\beta \lra
i\beta$ in (\ref{KB1}).
We find
\be\label{con}
K_{B_1^{(1)}}(\mu) \lra K_1(\th) = \bl -n \br\, \bl n+1-\frac B2
\br\, \bl -1+\frac B2 \br\;,
\end{equation}
in which we now use the block notation (\ref{bracketre}) and
the parameter $B=-2/\lambda$.
In analogy to the construction of the exact \mbox{$S$-matrices}
\cite{brade90}, it is possible to construct the reflection amplitudes for
all other particles from this lowest reflection amplitude. In $\ao$ ATFT
two particles, say $a$ and $b$, can fuse into a particle $a+b$ at the
rapidity difference $\th = i\pi\frac{a+b}{n+1}$ (if $a+b <n+1$). The
bootstrap principle for an integrable theory on the half-line states
that the reflection of two particles is independent of whether the
fusion into a third particle occurs before or after they reflect
off
the boundary. This leads to the following boundary bootstrap equation
\cite{fring94}
which is illustrated in figure~\ref{figbootstrap}:
\be
K_{a+b}(\th) = K_b\left(\th-i\pi\frac{a}{n+1}\right)\,
S_{a,b}\left(2\th+i\pi\frac{b-a}{n+1}\right)\,
K_a\left(\th+i\pi\frac{b}{n+1}\right)\;, \label{boundbootab}
\end{equation}
in which $S_{a,b}(\th)$ is the real coupling $\ao$ affine Toda
$S$-matrix, which can be written in the form
\be
S_{a,b}(\th) = \prod_{\stackrel{a-b+1}{\text{step }2}}^{a+b-1}
\bl p+1\br\, \bl p-1 \br\, \bl
-p-1+B \br\, \bl -p+1-B \br\;. \label{anSmatrix}
\end{equation}
We have chosen $a\leq b$.
\vs{0.2cm}
We can now obtain any $K_a(\th)$ from $K_1(\th)$ by applying
(\ref{boundbootab}) successively $a-1$ times:
\be
K_a(\th) = K_1\(\th+i\pi\frac{a-1}{n+1}\)\, \prod_{c=1}^{a-1}
K_1\(\th-i\pi\frac{a+1-2c}{n+1}\)\,
S_{a-c,1}\(2\th+i\pi\frac{3c-a-1}{n+1}\)\;. \label{KSK}
\end{equation}
Using the formula (\ref{anSmatrix}) for the $S$-matrix
we can compute the right hand side of (\ref{KSK}) explicitly, and we
obtain
\begin{equation}
K_a(\th) = \prod_{c=1}^a \bl c-1 \br\, \bl c-n-1 \br\, \bl -c+\frac
B2 \br\, \bl -c-n-\frac B2 \br\;. \label{Ka}
\end{equation}
It is interesting to note that this result agrees with a
conjecture which was made for these
reflection amplitudes already 5 years ago in \cite{corri94}.
It is easy to check that these reflection amplitudes satisfy
the necessary boundary crossing relation
\begin{equation}
K_a(\th)K_{\bar a}(\th+i\pi)=S_{aa}(2\th).
\end{equation}
We also find that the amplitudes are invariant under
charge-conjugation,
\be
K_a(\th) = K_{\bar a}(\th)\;.
\end{equation}
The classical limit ($B\to 0$) of $K_a(\th)$ is
\be
K_a(\th) \lra -\bl -a \br\, \bl a+n+1 \br\;.
\end{equation}
This agrees with the classical result in formula (4.22) of \cite{bowco96b}.

In the sine--Gordon theory \cite{corri97} as well as the $\ato$ ATFT
\cite{gande98b} it was found
that the positive boundary condition and the Neumann boundary condition
are dual to each other under the weak--strong coupling duality $\beta
\lra \frac{4\pi}{\beta}$. Therefore, we conjecture that the dual
reflection amplitude to (\ref{Ka}) describes the reflection of $\ao$
affine Toda particles from a boundary with Neumann boundary
condition.
The dual of (\ref{Ka}) can simply be obtained by changing $B \to
2-B$,
\be
K_a(\th) \lra K^N_a(\th) = \prod_{c=1}^a \bl c-1 \br\, \bl
c-n-1 \br\, \bl -c+1 - \frac B2 \br\, \bl -c+n+1 +\frac B2 \br\;.
\label{KNeum}
\end{equation}
First we check the
classical limit of these amplitudes and we find
\be
K^N(\mu) \lra 1\;, \;\;\;\;\;\;\;\; (B \to 0)\;,
\end{equation}
which supports the conjecture that $K_a^N(\mu)$ are the reflection
amplitudes for the Neumann boundary condition.
In fact, here we can go one step further since Kim \cite{kim95b} has
computed the reflection amplitudes for the Neumann boundary condition up
to first order in the coupling constant.
If we expand $K^N(\mu)$ in terms of $\beta^2$ and compare the
${\cal O}(\beta^2)$ term with the perturbative result of \cite{kim95b}
we find that they are indeed identical.

\subsection{CDD pole ambiguities\label{scdd}}
The scalar factor in the soliton reflection
matrix is not uniquely fixed by the equations \eqref{aeq}.
Rather the minimal solution given in \eq{aa}
can be multiplied by any factor
$\sigma(\mu)$ which satisfies
\bea
\si(\mu) \si(-\mu) &=& 1 \;, \nn \\
\si(-\mu +\frac{n+1}4 \la) &=& \si(\mu+\frac{n+1} 4 \la) \;.
\eea
Such ambiguities are often referred to as CDD pole ambiguities.
In order for the soliton bootstrap equations to be satisfied the
factors must obey the additional condition
\begin{equation}\label{ext}
\si(\mu) = \prod_{k=1}^n \si\left(\mu+\frac{n+1}2 \la - k\la \right)\;.
\end{equation}
These equations are solved by the following expressions
\begin{equation}\label{cdd}
\si_c(\mu) = - \bl \frac c2 \la \br_I \, \bl \frac{n+1}2 \la -\frac c2
\la \br_I\;,
\end{equation}
for $c= 1,2,\dots \leq \frac{n+1}2$ (actually $c$ can run from $1$ to
$n$ but that does not give any new factors since $\si_c(\mu) =
\si_{n+1-c}(\mu)$.)
The inverse of this, i.e. $\( \si_c(\mu) \) ^{-1} = - \bl
- \frac c2 \la \br \, \bl \frac c2 \la - \frac{n+1}2 \la \br $, is also
a solution of \eq{ext}.
The possible CDD factors are products of any of these
blocks.

After the bootstrap to the breather reflection amplitude and
analytic continuation to the particle amplitude a
soliton CDD factor $\si_c(\mu)$ would lead to a particle CDD factor
\begin{equation}
k_c(\mu) = \bl \frac{n+1}2 +c + \frac B2 \br\, \bl \frac{n+1}2 -c - \frac
B2 \br\, \bl - \frac{n+1}2 +c - \frac B2 \br\,\bl -\frac{n+1}2 - c +
\frac B2 \br\;.
\end{equation}
The classical limit ($B \to 0$) of these factors is equal to $1$.

We suspect that it will be
impossible to find a physical explanation of the extra poles
supplied by these CDD factors. Therefore we will from now on work
with the solution without CDD factors for which we will show the
consistency of the pole structure in section \ref{sect:poles}.

\section{The vacuum solutions for the solitonic boundary conditions
\label{sect:vacuum}}

We will now discuss the vacuum solutions to real
coupling $a_n^{(1)}$ affine Toda theory on the left half-line
with boundary condition of the form \eqref{bcond},
\be\label{bcond2}
\left.\beta\,\partial_x\bphi+\sum_{i= 0}^n \bp_i\,\aroot_i
e^{\beta\aroot_i\cdot\bphi/2}\right|_{x= 0}= 0
\end{equation}
Two of these boundary conditions are rather special because their
vacuum solution is simply the constant solution
$\bphi= 0$. The simplest is the Neumann condition which is obtained
if all the boundary parameters $\bp_i$ are equal to zero. The
other is the boundary condition which has all $\bp_i$ equal to
$+1$. We will refer to it as the uniform $(++\cdots++)$ boundary condition.
The boundary condition which has all $\bp_i$ equal to
$-1$ is also satisfied by the constant solution $\bphi=0$ but for
this boundary condition there exists another real solution with
the same energy.

In \cite{deliu98} we explained that if the $\bp_i$ are equal
to $\pm 1$ with at least one being $-1$ and if they satisfy the
constraint
\be\label{bpcond}
\prod_{i= 0}^n\bp_i= 1,
\end{equation}
then the vacuum solution is
obtained by taking a solution describing a stationary soliton
on the left half-line and continuing it to real $\beta$.
The soliton has to have a topological charge $\weight$ determined
by
\be\label{bptop}
(-1)^{\aroot_i\cdot\weight}= \bp_i,~~~~~~\forall i= 1,\dots,n.
\end{equation}
We will refer to these boundary conditions as the solitonic boundary
conditions.

For details of how we put a stationary soliton in front of the
boundary we refer to \cite{deliu98}. It turns out that a soliton
with topological charge $\weight$ gives the same vacuum solution as
a soliton with topological charge $-\weight$, which is good
because both $\weight$ and $-\weight$ give the same boundary
condition if substituted into \eq{bptop}.
In this section we will
explain how to find this topological charge $\weight$ given
the boundary parameters $\bp_i$.

It is known that the topological charges of the single soliton
solutions all lie in fundamental representations of $a_n$.
A fundamental representation $\Lambda_i$ is the representation whose
highest weight is the fundamental weight $\weight_i$.
The fundamental weights $\weight_i$ are defined by their
property
\[
\weight_i\cdot \aroot_j= \delta_{ij},~~~~~i,j= 1,\dots,n.
\]

Any $\weight$ can be written as a linear combination of the
fundamental weights, $\weight= \sum_{i= 1}^n a_i\weight_i$. It is
clear
that \eq{bptop} implies that the coefficient $a_i$ is even
if $\bp_i= 1$ and it is odd if $\bp_i= -1$. We will
show below that the unique positive weight
$\weight$ which satisfies \eq{bptop} and which lies
in a fundamental representation is obtained by setting all
the even $a_i$ to zero and letting the odd $a_i$ alternate between
$+1$ and $-1$ (with the highest non-vanishing coefficient being
$+1$ so that the weight is positive).
Furthermore one can determine the particular
fundamental representation $\Lambda_a$ in which the weight $\weight$
lies by adding
the indices of the fundamental weights multiplied by their
coefficient in the expansion of $\weight$. We illustrate this rule by
a few examples:
\begin{align}
{\bf \bp}&= (-1,-1)&
\weight&= \weight_2-\weight_1
\in\Lambda_i&\text{with }&i= 2-1= 1,\\
{\bf \bp}&= (1,-1,1,-1)&
\weight&= \weight_4-\weight_2
\in\Lambda_i&\text{with }&i= 4-2= 2,\\
{\bf \bp}&= (-1,-1,1,1,-1)&
\weight&= \weight_5-\weight_2+\weight_1
\in\Lambda_i&\text{with }&i= 5-2+1= 4,
\end{align}
where ${\bf \bp}= (\bp_1,\bp_2,\dots,\bp_n)$. $\bp_0$ is not needed
because it is determined in terms of the other $\bp_i$'s by
the condition in \eqref{bpcond}, namely $\bp_0= \prod_{i= 1}^n\bp_i$.

Let us summarise the above rule for the relation between the
boundary parameters $\bp_i$ and the topological charge
$\lambda$ of the soliton in a theorem.
\begin{theorem}\label{t1}
Let $C_i$ be the parameters specifying a solitonic boundary.
Let $P= \{k_1,k_2,\dots,k_a\}$ be the ordered set of indices
(with $k_1<k_2<\dots<k_a$) for which the boundary parameters
are $-1$, i.e., $\bp_k= -1$ for $k\in P$.
If $P$ is not empty then there is a unique positive weight
$\weight$ which satisfies
\be\label{e46}
(-1)^{\aroot_i\cdot\weight}= \bp_i,~~~~~~\forall i= 1,\dots,n,
\end{equation}
and lies in a fundamental representation. It is given by
\be\label{e47}
\weight= \sum_{j= 1}^a(-1)^{a-j}\weight_{k_j},
\end{equation}
and lies in the $i$-th fundamental representation where
\be\label{e48}
i= \sum_{j= 1}^a(-1)^{a-j}k_j.
\end{equation}
\end{theorem}
The rest of this section is devoted to the proof of this theorem
and can be skipped.
\vspace{3mm}
We needed one elementary fact
from the representation theory of simple complex Lie algebras:
\begin{theorem}\label{wt}
If $\weight$ is a weight of some irreducible representation
then $\weight-\aroot_i$ is also a weight of the representation
iff $\weight\cdot\aroot_i>0$. All weights of the representation
are obtained from the highest weight by repeatedly subtracting simple
roots in all possible
ways according to this rule.
\end{theorem}
We will use this to construct the weights of the fundamental
representations of $a_n$. It is convenient to introduce
an $n+1$ dimensional euclidean space with orthonormal basis vectors
$\e_i,~ i= 1,\dots,n+1$. We embed the $n$ dimensional root space of
$a_n$ into this space such that it is the
hyperplane perpendicular to the vector
$\e= \sum_{i= 1}^{n+1}\,\e_i$. An overcomplete and non-orthonormal
(but nevertheless useful)
basis for the root space is given by the
$n+1$ vectors
\be
\er_i= \e_i-\frac{1}{n+1}\,\e,~~~~~i= 1,\cdots,n+1.
\end{equation}
It is overcomplete because $\er_{n+1}= -\sum_{i= 1}^n\er_i$.
The simple roots can be chosen to be
\be
\aroot_i= \e_i-\e_{i+1}= \er_i-\er_{i+1},~~~~~i= 1,\dots,n.
\end{equation}
The corresponding fundamental weights are
\be\label{fw}
\weight_i= \sum_{a= 1}^i\er_a,~~~~~i= 1,\dots,n.
\end{equation}
We can now give the expressions for all the weights in the
fundamental representations.
\begin{proposition}
The weights of the $i$-th fundamental representation $\Lambda_i$
are all vectors that are given as a sum of $i$ different basis
vectors,
\be
\Lambda_i= \left\{\left.\sum_{j= 1}^i\er_{k_j}\right|
k_j\neq k_{j'} \text{ for }j\neq j'\right\}.
\end{equation}
In other words, the weight vectors of $\Lambda_i$ are all vectors
with $i$ $1$'s and $n-i$ $0$'s when expanded in the $\er$ basis.
They are positive if the $n+1$st entry is 0.
The dimension of the $i$-th fundamental representation is
\be
\rm{dim}\, \Lambda_i= \begin{pmatrix}n+1\\i\end{pmatrix}.
\end{equation}
\end{proposition}
\begin{proof}
Consider any weight vector $\weight$
whose entries are only $1$'s and $0$'s. The scalar
product of such a vector with a simple root $\aroot_j$ is positive
only if the vector has a $1$ at the $j$-th position and a
$0$ at the $j+1$-st position. Then and only then will $\weight-\aroot_j$
also be a weight, according to theorem \ref{wt}. The effect of
subtracting
$\aroot_j$ from $\weight$ will be to shift the $1$ at the $j$-th
position
one step to the right. Doing this repeatedly one can transform the
highest weight vector $\lambda_i= (1,\dots,1,0,\dots,0)$,
which has all the $i$ $1$'s to the left and the $0$'s
to the right, into any other vector with $i$ $1$'s. Furthermore one
will never obtain a weight vector with more or less than $i$ $1$'s.
The number of different vectors with $i$ $1$'s and the rest $0$'s is
equal to
the number of ways one can choose $i$ sites out of $n+1$ and this gives
the dimension of the $i$-th fundamental representation.
\end{proof}

If one inverts the expressions \eqref{fw}
for the fundamental weights
one arrives at the following expressions for the basis vectors in
terms of the fundamental weights
\be
\er_1= \weight_1,~~~~
\er_i= \weight_i-\weight_{i-1},~~~i= 2,\dots,n,~~~~\er_{n+1}=
-\weight_n.
\end{equation}
Substituting these in the above proposition one obtains the following
corollary.

\begin{corollary}\label{cor}
The positive weights in the $i$-th fundamental representation are linear
combinations of fundamental weights,
$\lambda= \sum_{k= 1}^na_k\weight_k$,
with the following restrictions on the
coefficients:
\begin{enumerate}
\item The coefficient of
the largest occurring fundamental weight is $1$, the coefficient of
the next largest occurring fundamental weight is $-1$ and so on,
the non-zero coefficients alternating between $+1$ and $-1$.
\item The coefficients multiplied by the indices of
the corresponding fundamental weights sum up to $i$,
\be
\sum_{k= 1}^n k\,a_k= i.
\end{equation}
\end{enumerate}
\end{corollary}

We are now in a position to prove our main theorem \ref{t1}.
The fact that a weight $\weight$ given by \eq{e47} lies in the $i$-th
fundamental representation with $i$ given by \eq{e48} is simply a
rewriting of Corollary \ref{cor}. It is obvious from
$\weight_i\cdot\aroot_j=\delta_{ij}$ that this $\weight$ satisfies
\eq{e46}. To
prove that $\weight$ is the unique positive weight in a
fundamental representation satisfying \eq{e46} we only have to
observe that
the total number of weights in the fundamental representations is
\be
\sum_{i= 1}^n \text{dim}\Lambda_i= \sum_{i= 1}^n
\begin{pmatrix}n+1\\i\end{pmatrix}= 2^{n+1}-2
\end{equation}
and thus the number of positive weights is equal to $2^n-1$ which is
also equal to the number of boundary conditions which satisfy
\eq{bpcond} and don't have all $\bp_i$ equal to $1$.

\section{The reflection amplitudes for the solitonic boundary conditions
\label{sect:mixed}}

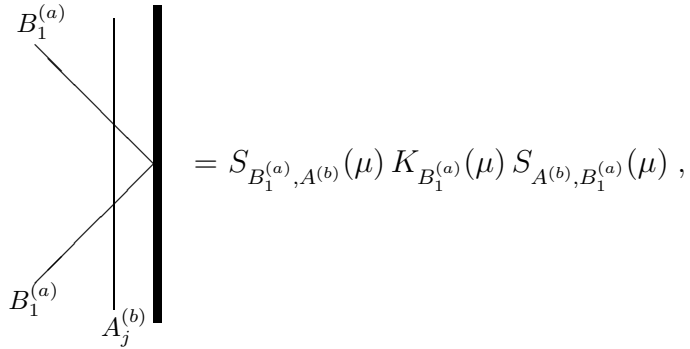
\begin{figure}[ht]
\begin{center}
\begin{picture}(150,140)(-10,0)
\put(50,10){\rule{3pt}{120pt}}
\put(50,70){\line(-1,-1){45}}
\put(50,70){\line(-1,1){45}}
\put(35,15){\line(0,1){110}}
\put(65,67){\shs{ = $S_{B^{(a)}_1,A^{(b)}}(\mu)\, K_{B^{(a)}_1}(\mu)\,
S_{A^{(b)},B^{(a)}_1}(\mu)\;,$}}
\put(-5,15){\shs{\fns{$B^{(a)}_1$}}}
\put(-2,120){\shs{\fns{$B^{(a)}_1$}}}
\put(30,5){\shs{\fns{$A^{(b)}_j$}}}
\end{picture}
\end{center}
\caption{\it  A stationary soliton in front of the
boundary}\label{figstatsol}
\end{figure}

We can now utilise the result from the preceding section in order to
construct the reflection amplitudes for the solitonic boundary conditions.
We have seen that we should be able to obtain them
by putting a stationary soliton in front of the boundary. In the case
of an incoming breather this leads to an additional breather--soliton scattering
process before and after the reflection off the boundary as illustrated
in figure~\ref{figstatsol}. The new reflection amplitude is then
just the product of the two $S$-matrix elements and the reflection amplitude
for the uniform $(++\cdots ++)$ boundary.

The scattering amplitudes of a soliton with a breather in imaginary
coupling $\ao$ ATFT was given in \cite{gande98}. There it was found
that this type of scattering process is independent of the
multiplet label $j$ of the soliton, and the scattering amplitudes were
found to be (using the bracket notation defined in \eq{bracketim})
\be
S_{A^{(b)},B_1^{(a)}}(\mu) = \prod_{k=1}^a \frac{
\lb \frac{\la}2 (a+b-\frac{n+1}2 -2k+2) +\frac 12 \rb\,
\lb \frac{\la}2 (a-b-\frac{n+1}2 -2k) -\frac 12 \rb}
{\lb \frac{\la}2 (a-b-\frac{n+1}2 -2k+2) +\frac 12 \rb\,
\lb \frac{\la}2 (a+b-\frac{n+1}2 -2k) - \frac 12 \rb}\;,
\end{equation}
and
\be
S_{B^{(a)}_1,A^{(b)}}(\mu) =
S_{A^{(n+1-b)},B_1^{(a)}}(\mu)\;. \label{BAsymm}
\end{equation}
Therefore, the reflection process in figure~\ref{figstatsol}
can be computed and we find
\bea
\lefteqn{S_{B^{(a)}_1,A^{(b)}}(\mu)\, K_{B_1^{(a)}}(\mu)\,
S_{A^{(b)},B^{(a)}_1}(\mu) =} \hs{1cm} \nn \\
&& = K_{B_1^{(a)}}(\mu)\,\prod_{k=1}^a
\frac{\bl \frac{\la}2 (a+b+\frac{n+1}2 -2k) -\frac 12 \bri\,
\bl \frac{\la}2 (b-a-\frac{n+1}2 +2k) +\frac 12 \bri}
{\bl \frac{\la}2 (a+b-\frac{n+1}2 -2k) -\frac 12 \bri\,
\bl \frac{\la}2 (b-a+\frac{n+1}2 +2k) + \frac 12 \bri} \;,
\label{addamplitude}
\end{eqnarray}
in which $K_{B_1^{(a)}}(\mu)$ is the amplitude describing the reflection
of the breather $B_1^{(a)}$ from the (++...+) boundary.

After analytic continuation the additional factor from the
two soliton--breather scattering amplitudes in \eq{addamplitude}
becomes
\be
\kextra^{(b)}_a(\th)
= \prod_{k=1}^a  \frac{\bl a+b+\frac{n+1}2 -2k +\frac B2 \br\,
\bl b-a-\frac{n+1}2 +2k - \frac B2 \br}
{\bl a+b-\frac{n+1}2 -2k + \frac B2 \br\,
\bl b-a+\frac{n+1}2 +2k - \frac B2 \br}\;.
\label{statsolfactor}
\end{equation}
We thus obtain the following reflection amplitudes for the particles in $\ao$ ATFT
with solitonic boundary conditions:
\be
K^{(b)}_a(\th) = K_a(\th)\, \kextra^{(b)}_a(\th)\;, \label{Kmixed}
\end{equation}
in which $K_a(\th)$ is given by (\ref{Ka}).
Because of the symmetry relation (\ref{BAsymm}) we find that
\be
K^{(b)}_a(\th) = K^{(n+1-b)}_a(\th)\,.
\end{equation}
This is needed for consistency because if a soliton of weight
$\weight$ lies in representation $b$ then its antisoliton of
weight $-\lambda$ lies in the conjugate representation $n+1-b$ and
as we had mentioned in section \ref{sect:vacuum}, both of these solitons
give rise to the same vacuum solution and should therefore also
give the same particle reflection factors.

Something that wasn't obvious from the classical calculations is
the fact that all solitonic boundary conditions corresponding to
topological charges $\weight$ in the same fundamental
representation $V_b$ or its dual $V_{n+1-b}$
lead to the same reflection amplitudes.
There are only
$\left\lfloor\frac{n+1}{2}\right\rfloor$ classes of boundary conditions,
each corresponding to a pair of conjugate representations.
The classical vacuum solutions
look quite different for different members of these classes but their
quantum reflection amplitudes turn out to be the same.

It is again easy to check that the above reflection amplitudes
(\ref{Kmixed}) satisfy the necessary boundary crossing and
unitarity conditions.
We can also look at the classical limit of these amplitudes and we find
\bea
\lefteqn{\kextra^{(a)}_b(\mu) \lra   \prod_{k=1}^b  \frac{\bl
a+b+\frac{n+1}2 -2k \br\, \bl a-b-\frac{n+1}2 +2k \br} {\bl
a+b-\frac{n+1}2 -2k \br\,
\bl a-b+\frac{n+1}2 +2k \br}} \hs{0.5cm} \nn \\
&&= \bl a+b-\frac{n+1}2\br \, \bl -a-b-\frac{n+1}2 \br \, \bl
a-b+\frac{n+1}2\br \, \bl b-a+\frac{n+1}2\br\,,
\end{eqnarray}
which turns out to be the same as formula (4.21) in the paper
\cite{bowco96b}, whereas the dual of $\kextra^{(a)}_b(\mu)$ has again
a classical limit of $1$.

\section{Poles and boundary bound states
\label{sect:poles}}

In order to show the consistency of all reflection amplitudes we have
constructed, we need to examine their pole structure and
show that all physical strip poles can be explained in terms of
boundary bound states or in terms of some other processes which can
be regarded as analogues of the generalised Coleman-Thun mechanism in
the bulk theory \cite{dorey98}.

\subsection{Fixed poles\label{fixed}}

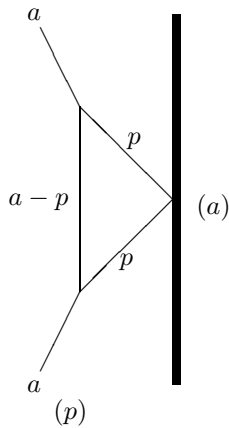
\begin{figure}[ht]
%
%
\begin{center}
\begin{picture}(200,170)(0,0)
\put(120,10){\rule{3pt}{140pt}}
\put(85,115){\line(-1,2){15}}
\put(85,45){\line(0,1){70}}
\put(120,80){\line(-1,-1){35}}
\put(120,80){\line(-1,1){35}}
\put(85,45){\line(-1,-2){15}}
\put(65,7){\shs{\fns{$a$}}}
\put(100,55){\shs{\fns{$p$}}}
\put(103,101){\shs{\fns{$p$}}}
\put(58,78){\shs{\fns{$a-p$}}}
\put(65,148){\shs{\fns{$a$}}}
\put(75,-3){\shs{\fns{$\left(p\right)$}}}
\put(129.00,77.00){\makebox(0,0)[lc]{\shs{\fns{$\left(a\right)$}}}}
\end{picture}
\end{center}
\caption{\it On-shell triangle diagram producing the pole at $\th=\frac{i\pi}{n+1}p$
in the reflection amplitude $K_a$ with $a\leq\frac{n+1}{2}$.\label{figtriangle}}
\end{figure}

All $\ao$ affine Toda particle reflection amplitudes, as given in
(\ref{Ka}, \ref{KNeum}, \ref{Kmixed}), contain the same coupling constant
independent part and we can therefore discuss the pole structure in
this part collectively for all boundary conditions. Because
$K_a=K_{n+1-a}$ we only need to consider the case
$a\leq\frac{n+1}{2}$. Once
we have found the diagrams which explain the poles in $K_a$, the
poles in $K_{n+1-a}$ are explained by the charge conjugated
versions
of the same diagrams (i.e., we simply have to replace every
particle in a diagram by its antiparticle).
The term $\prod_{c=1}^a \bl c-1 \br\, \bl c-n-1 \br$ contains
poles on the physical strip at
\be
\th = i\pi \frac{p}{n+1}\;, \hs{1.5cm} p = 1,2,\dots,
a-1\;.
\end{equation}
The location of these poles does not depend on the
coupling constant.
They arise from the triangle diagram depicted in figure~\ref{figtriangle}
in which all lines are on-shell.
It is clear that this reflection process only exists for $p=1,2,\dots
a-1$ because otherwise particle $a-p$ does not exist.

The angles in the diagram are not drawn to scale. They are
determined by the requirement that the internal particle lines
should be on shell. The angles between the particle lines
and the vertical are equal to $-i$ times the rapidity of the
particles.
Thus, for example, to determine the angle
between the lines of particles $p$ and $a-p$ we only need to know
the relative rapidity at which particles $p$ and $a-p$ will fuse
to form the particle $a$. This is $\th=\frac{i\pi}{n+1}a$, as can
be read off from either the masses of the particles or from the
location of the corresponding pole in the S-matrix element
$S_{p,a-p}$. The angle is therefore equal to
$\frac{\pi}{n+1}a$. This is also the angle between the line of
particle $p$ and the boundary. In diagram \ref{figtriangle}, as in
all diagrams which follow, we have indicated the angle at which a
particle line meets the boundary by giving a number in parentheses
which states the angle as a multiple of $\frac{\pi}{n+1}$.
Similarly, we give the angle of the incoming line in the lower left
hand of the diagram. This gives the location of the pole produced
by the diagram.

Diagram \ref{figtriangle} contains three internal lines and
one loop and will therefore lead to a simple pole in the reflection
amplitude for particle $a$, provided that the reflection of
particle $p$ does not take place at a rapidity at which $K_p(\th)$
displays a pole or a zero. The rapidity of particle $p$ is
$\th=\frac{i\pi}{n+1}a$, as can be
read off from the angle given. Because $p<a$, $K_p$ does not have
any pole or zero at this location and therefore the diagram
does indeed lead to a simple pole.

This diagram explains all $\beta$ independent poles in all $\ao$ reflection
amplitudes. In the Neumann as well as the (++...+) boundary reflection
amplitudes there are no further poles. However, for the reflection
amplitudes (\ref{Kmixed}) for the solitonic boundaries
there are coupling constant dependent poles. Below we will study
the pole structure in detail for $a_2^{(1)}$ and $a_4^{(1)}$ Toda
theories. We will restrict our attention to a coupling constant in
the range $0<B<1$ (recall that in general $0\leq B\leq 2$) and we
will exclude any non-generic rational values at which special
cancellations might take place.

\subsection{$\ato$ ATFT\label{subsect:mixed2}}

In this subsection we have
\be
\bl a \br \equiv \frac{\sin(\frac{\th}{2i} +\frac{a\pi}{6})}
{\sin(\frac{\th}{2i} -\frac{a\pi}{6})}\;.
\end{equation}
In the case of $\ato$ there are three solitonic boundary conditions,
$(+ - -)$, $(- + -)$ and $(- - +)$.
Their vacuum solutions are given by placing the appropriate stationary
soliton from the vector representation in front of the boundary.
They all lead to the same reflection amplitude
\be
K^{b_{0,0}}_1(\th) = K^{b_{0,0}}_2(\th) = K_1(\th) \bl \frac 12 - \frac B2
\br\, \bl
\frac 32 - \frac B2 \br\, \bl \frac 32 + \frac B2 \br\, \bl \frac 52 +
\frac B2 \br\;, \label{a2K}
\end{equation}
in which
\be\label{kkk}
K_1(\th) =\bl -2 \br\,  \bl -1 +\frac B2 \br\,  \bl 3 -\frac B2
\br\;= K_2(\th)\;.
\end{equation}
We have labelled the vacuum state by $b_{0,0}$ to distinguish it from
the excited boundary states $b_{n,m}$ to be derived
below.
In \cite{corri94} a calculation of the classical reflection
amplitude had lead to an expression for this reflection amplitude
containing an unknown function $C$ of the coupling constant. The conjecture
coincides with our result if one sets $C=-\frac{B}{2}$.

$K_1$ contains no poles in the physical strip. So there are two
physical strip poles in (\ref{a2K}), namely at
\be
\th_l = \frac{i\pi}3\left(\frac 32 - \frac B2\right)\;, \hs{1cm} \mbox{and}\;\;\;
\th_h = \frac{i\pi}3\left(\frac 12 - \frac B2\right)\;,
\end{equation}
where $\th_h$ only exists in the physical strip if $B\leq 1$.
We restrict our analysis to this case.

Let us begin with the pole $\th_h$ in $K_1^{b_{0,0}}$, which we expect to
correspond to a boundary bound
state (or excited boundary). The reflection amplitudes for the
reflection from this boundary bound state can easily be computed
using the boundary bootstrap equation \eq{bb}.
The boundary bound state formed by an incoming particle 1 with
rapidity $\th_h$ will be denoted by $b_{1,0}$.
The reflection amplitude for particle 1 scattering off boundary
$b_{1,0}$ is
\bea
K_1^{b_{1,0}}(\th) &=& K_1(\th) \bl \frac 12 -\frac B2 \br\, \bl
-\frac 12 -\frac B2
\br\, \bl \frac 32 +\frac B2 \br^2\nn\\
&& \hs{30pt} \times \bl \frac 52 -\frac B2 \br\, \bl
\frac 52 +\frac B2 \br\,
\bl \frac 12 -\frac 32 B \br\,
\bl -\frac 52
+\frac 32 B \br\;.
\end{eqnarray}
We find that this reflection amplitude again contains physical strip
poles, some of which correspond to new boundary bound
states. Through repeated application of the boundary bootstrap
equation and by considering all possible physical strip poles, we
eventually obtain a whole array of boundary bound states, which
can be labelled by $b_{n,m}$, in which $n$ ($m$) is the number of
particles of type 1 (2) bound to the boundary.
In addition it turns out that $n$ or $m$ can be negative,
but only as long as $n+m \geq 0$. This ensures that there are no
boundary states with energies lower than the ground state
$b_{0,0}$.

The general reflection amplitude for a particle 1 scattering off a
boundary $b_{n,m}$ can be computed and we find
\bea
K_1^{b_{n,m}}(\th) &=& K_1(\th)\, \bl \frac 12 - (2n+1)\frac B2
\br \, \bl \frac 52 + (2m+1)\frac B2 \br \nn \\
&& \hs{31pt} \times   \bl \frac 12 - (2n-1)\frac B2
\br \, \bl \frac 52 + (2m-1)\frac B2 \br \nn \\
&& \hs{31pt} \times   \bl \frac 32 + (2n-1)\frac B2
\br \, \bl \frac 32 - (2m-1)\frac B2 \br \nn \\
&& \hs{31pt} \times   \bl -\frac 52 + (2n+1)\frac B2
\br \, \bl -\frac 12 - (2m+1)\frac B2 \br\;. \label{Khnm}
\end{eqnarray}
It is easy to check that these  reflection amplitudes all satisfy the
necessary boundary unitarity, crossing and bootstrap conditions.
The blocks in (\ref{Khnm}) have been arranged such that the second
block in each line is the crossing transform of the first block in
this line, i.e.\ each line is of the form $\bl a(n,m) \br \, \bl h -
a(m,n) \br$.
The reflection amplitudes for particle 2 can be obtained by
charge conjugation
\be
K_2^{b_{n,m}}(\th) = K_1^{b_{m,n}}(\th)\;.
\end{equation}

In order to test the consistency of these reflection amplitudes, we now
have to explain all the poles in these new reflection amplitudes.
To do this we will need to know the possible particle fusings for
$a_2^{(1)}$. They are $1+1\rightarrow 2$ and $2+2\rightarrow 1$,
both occurring at a relative rapidity of $\th=2\frac{i\pi}{3}$.

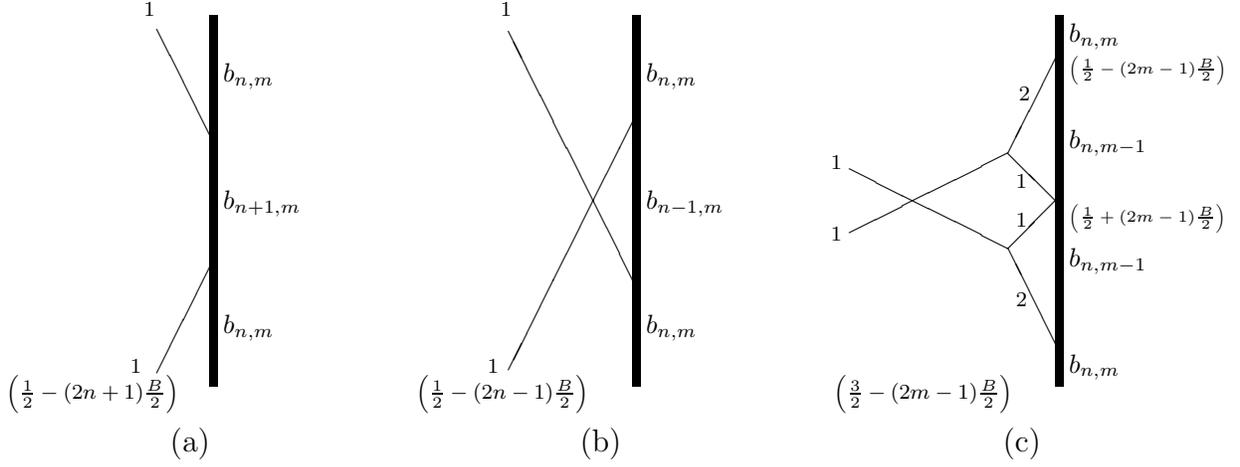
\begin{figure}[ht]
\begin{picture}(480,190)(30,-10)
\put(120,10){\rule{3pt}{140pt}}
\put(120,55){\line(-1,-2){20}}
\put(120,105){\line(-1,2){20}}
\put(90,15){\shs{\scs{$1$}}}
\put(95,150){\shs{\scs{$1$}}}
\put(125,30){\shs{\fns{$b_{n,m}$}}}
\put(125,77){\shs{\fns{$b_{n+1,m}$}}}
\put(125,125){\shs{\fns{$b_{n,m}$}}}
%
%
\put(280,10){\rule{3pt}{140pt}}
\put(280,110){\line(-1,-2){47}}
\put(280,50){\line(-1,2){47}}
\put(225,15){\shs{\scs{$1$}}}
\put(230,150){\shs{\scs{$1$}}}
\put(285,30){\shs{\fns{$b_{n,m}$}}}
\put(285,77){\shs{\fns{$b_{n-1,m}$}}}
\put(285,125){\shs{\fns{$b_{n,m}$}}}
%
%
\put(440,10){\rule{3pt}{140pt}}
\put(440,26){\line(-1,2){18}}
\put(440,134){\line(-1,-2){18}}
\put(440,80){\line(-1,-1){18}}
\put(440,80){\line(-1,1){18}}
\put(422,62){\line(-2,1){60}}
\put(422,98){\line(-2,-1){60}}
\put(355,65){\shs{\scs{$1$}}}
\put(355,92){\shs{\scs{$1$}}}
\put(425,40){\shs{\scs{$2$}}}
\put(426,118){\shs{\scs{$2$}}}
\put(425,70){\shs{\scs{$1$}}}
\put(425,85){\shs{\scs{$1$}}}
\put(445,15){\shs{\fns{$b_{n,m}$}}}
\put(445,55){\shs{\fns{$b_{n,m-1}$}}}
\put(445,100){\shs{\fns{$b_{n,m-1}$}}}
\put(445,140){\shs{\fns{$b_{n,m}$}}}
\put(445,128){\shs{\tiny{$\(\frac 12-(2m-1)\frac B2\)$}}}
\put(445,72){\shs{\tiny{$\(\frac 12+(2m-1)\frac B2\)$}}}
%
\put(43,5){\shs{\scs{$\(\frac 12 -(2n+1) \frac B2\)$}}}
\put(198,5){\shs{\scs{$\(\frac 12 -(2n-1)\frac B2\)$}}}
\put(356,5){\shs{\scs{$\(\frac 32 -(2m-1)\frac B2\)$}}}
\put(105,-15){\shs{ (a)}}
\put(260,-15){\shs{ (b)}}
\put(420,-15){\shs{ (c)}}
\end{picture}
\caption{\it Poles in $K_1^{b_{n,m}}$ with $n+m>0$.}\label{figa2pol1}
\end{figure}

\begin{figure}[ht]
\begin{center}
\begin{picture}(280,190)(0,-10)
\put(120,10){\rule{3pt}{140pt}}
\put(120,26){\line(-1,1){18}}
\put(120,134){\line(-1,-1){18}}
\put(120,80){\line(-1,-2){18}}
\put(120,80){\line(-1,2){18}}
\put(102,44){\line(-2,-1){60}}
\put(102,116){\line(-2,1){60}}
\put(35,10){\shs{\scs{$1$}}}
\put(35,143){\shs{\scs{$1$}}}
\put(108,28){\shs{\scs{$1$}}}
\put(108,127){\shs{\scs{$1$}}}
\put(106,61){\shs{\scs{$2$}}}
\put(106,94){\shs{\scs{$2$}}}
\put(125,15){\shs{\fns{$b_{n,m}$}}}
\put(125,55){\shs{\fns{$b_{n-1,m}$}}}
\put(125,100){\shs{\fns{$b_{n-1,m}$}}}
\put(125,140){\shs{\fns{$b_{n,m}$}}}
\put(45,0){\shs{\scs{$\(\frac 32 +(2n-1)\frac B2\)$}}}
\put(125,130){\shs{\tiny{$\(\frac 12-(2n-1)\frac B2\)$}}}
\put(125,72){\shs{\tiny{$\(\frac 12+(2n-1)\frac B2\)$}}}
%
%
\end{picture}
\end{center}
\caption{\it Extra pole in $K_1^{b_{n,m}}$ for
$n \leq 0$.}\label{figa2pol2}
\end{figure}
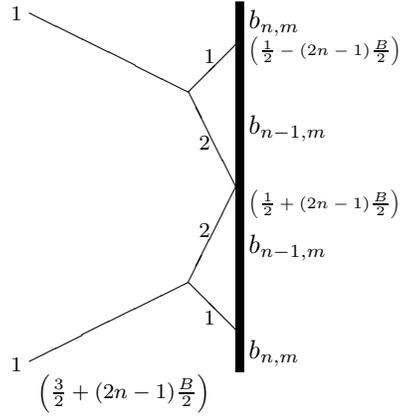

First let us consider the case of $n+m > 0$. There are physical strip
poles in the first blocks on the first and second line and in the
second block on the third line of \eq{Khnm}, which can be explained by the
diagrams in figure~\ref{figa2pol1}.
Note that diagram \ref{figa2pol1}(c) only exists if $m > 0$.
This agrees with the fact that the corresponding pole in the
amplitude moves off the physical strip for $m\leq 0$.
There is an additional pole on the physical strip
in the case of $n\leq 0$ (but still $n+m >0$)
in the first block in the third line. This pole can be explained in
terms of the reflection process in figure~\ref{figa2pol2}.
It can be checked that the reflections in the centre of diagrams
\ref{figa2pol1}(c) and \ref{figa2pol2} occur at a zero in the
corresponding reflection amplitudes, and
therefore these diagrams yield simple poles, as expected.

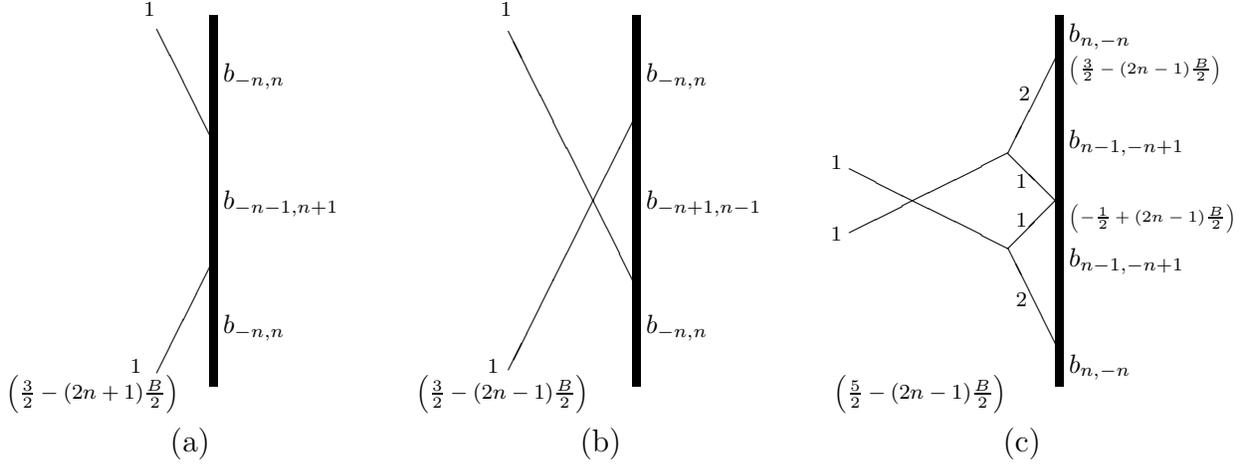
\begin{figure}[ht]
\begin{picture}(480,190)(30,-10)
\put(120,10){\rule{3pt}{140pt}}
\put(120,55){\line(-1,-2){20}}
\put(120,105){\line(-1,2){20}}
\put(90,15){\shs{\scs{$1$}}}
\put(95,150){\shs{\scs{$1$}}}
\put(125,30){\shs{\fns{$b_{-n,n}$}}}
\put(125,77){\shs{\fns{$b_{-n-1,n+1}$}}}
\put(125,125){\shs{\fns{$b_{-n,n}$}}}
%
%
\put(280,10){\rule{3pt}{140pt}}
\put(280,110){\line(-1,-2){47}}
\put(280,50){\line(-1,2){47}}
\put(225,15){\shs{\scs{$1$}}}
\put(230,150){\shs{\scs{$1$}}}
\put(285,30){\shs{\fns{$b_{-n,n}$}}}
\put(285,77){\shs{\fns{$b_{-n+1,n-1}$}}}
\put(285,125){\shs{\fns{$b_{-n,n}$}}}
%
%
\put(440,10){\rule{3pt}{140pt}}
\put(440,26){\line(-1,2){18}}
\put(440,134){\line(-1,-2){18}}
\put(440,80){\line(-1,-1){18}}
\put(440,80){\line(-1,1){18}}
\put(422,62){\line(-2,1){60}}
\put(422,98){\line(-2,-1){60}}
\put(355,65){\shs{\scs{$1$}}}
\put(355,92){\shs{\scs{$1$}}}
\put(425,40){\shs{\scs{$2$}}}
\put(426,118){\shs{\scs{$2$}}}
\put(425,70){\shs{\scs{$1$}}}
\put(425,85){\shs{\scs{$1$}}}
\put(445,15){\shs{\fns{$b_{n,-n}$}}}
\put(445,55){\shs{\fns{$b_{n-1,-n+1}$}}}
\put(445,100){\shs{\fns{$b_{n-1,-n+1}$}}}
\put(445,140){\shs{\fns{$b_{n,-n}$}}}
\put(445,128){\shs{\tiny{$\(\frac 32-(2n-1)\frac B2\)$}}}
\put(445,72){\shs{\tiny{$\(-\frac 12+(2n-1)\frac B2\)$}}}
%
\put(43,5){\shs{\scs{$\(\frac 32 -(2n+1) \frac B2\)$}}}
\put(198,5){\shs{\scs{$\(\frac 32 -(2n-1)\frac B2\)$}}}
\put(356,5){\shs{\scs{$\(\frac 52 -(2n-1)\frac B2\)$}}}
\put(105,-15){\shs{ (a)}}
\put(260,-15){\shs{ (b)}}
\put(420,-15){\shs{ (c)}}
\end{picture}
\caption{\it Poles in $K_1^{b_{-n,n}}$ and $K_1^{b_{n,-n}}$.}\label{figa2pol3}
\end{figure}

Now we turn to the case of $n+m = 0$ which is slightly different.
In this case the second and fourth line in (\ref{Khnm}) cancel
each other and we get
\bea
K_1^{b_{-n,n}}(\th) &=& K_1(\th)\, \bl \frac 12 + (2n-1)\frac B2
\br \, \bl \frac 52 + (2n+1)\frac B2 \br \nn \\
&& \hs{31pt} \times   \bl \frac 32 - (2n+1)\frac B2
\br \, \bl \frac 32 - (2n-1)\frac B2 \br \;. \label{Khn-n}
\end{eqnarray}
There are three possible poles in these amplitudes. The pole in $\bl \frac
12 + (2n-1)\frac B2 \br$ corresponds to a process \ref{figa2pol1}a) in
which the
boundary $b_{-n+1,n}$ is created. The poles in $\bl \frac 32 -
(2n+1)\frac B2 \br$ and in $\bl \frac 32 - (2n-1)\frac B2 \br$
correspond to the two processes in figure~\ref{figa2pol3}(a) and
b) respectively.

We also obtain
\bea
K_1^{b_{n,-n}}(\th) &=& K_1(\th)\, \bl \frac 12 - (2n+1)\frac B2
\br \, \bl \frac 52 - (2n-1)\frac B2 \br \nn \\
&& \hs{31pt} \times   \bl \frac 32 + (2n-1)\frac B2
\br \, \bl \frac 32 + (2n+1)\frac B2 \br \;. \label{Kh-nn}
\end{eqnarray}
Here, the pole $\bl \frac 12 - (2n+1)\frac B2 \br$ corresponds again to a
process of type \ref{figa2pol1}a) in which the boundary state
$b_{n+1,-n}$ is created.
Furthermore, if  $\frac 2{2n-1} \leq B \leq \frac 3{2n-1}$ there
is a physical strip pole in $\bl \frac 52 - (2n-1) \frac B2 \br$ which
can be explained by the process in diagram \ref{figa2pol3}(c).
It is easy to check that $K_1^{b_{n-1,-n+1}}(\th)$ displays a zero
at $\th=\frac{i\pi}{3}\left(-\frac 12+(2n-1)\frac B2\right)$
so that this process leads to a simple pole, as required.

For non-zero $B$ all the boundary bound state poles move outside of the
physical strip for large $n$ or $m$, and we therefore
obtain only a finite number of boundary bound states in the theory.
The following boundary bound states exist:
\bea\label{boundsa2}
&b_{n,m}\;,&\mbox{for all integers} \hs{5pt} n,m \hs{5pt} \mbox{with}
\hs{5pt} n+m > 0 \hs{5pt} \mbox{and} \hs{5pt}
-\frac 1{2B} - \frac 12 < n,m  < \frac 1{2B}+\frac 12\;, \nn \\
&b_{n,-n}  \hs{4pt} \mbox{and} \hs{4pt} b_{-n,n}\;, \hs{5pt}
&\mbox{for all integers} \hs{5pt} n \hs{5pt} \mbox{with}
\hs{5pt} 0\leq n < \frac 3{2B}+\frac 12\;. \nn
\eea
This spectrum is plotted in figures \ref{spectrum1} and
\ref{spectrum2} at two different values of $B$.

Note that a special phenomenon occurs whenever
$B=\frac{1}{2n}$ with $n$ integer. Then the states
$b_{-n-a,n+a}$ are degenerate with the states $b_{-n+a,n}$ for all
$a>0$ and similarly the states $b_{n+a,-n-a}$ are degenerate with
the states $b_{n,-n+a}$.
In the amplitude $K_1^{b_{-n,n}}$ given in \eq{Khn-n} the
blocks $\left(\frac 12+(2n-1)\frac B2\right)$ and
$\left(\frac 32-(2n+1)\frac B2\right)$ coincide at
$\left(1-\frac B2\right)$. One of them cancels against the block
$\left(-1+\frac B2\right)$ in $K_1$ (see \eq{kkk}).
This leaves
only a single pole which can receives a contribution both from
the propagation of
the boundary state $b_{-n+1,n}$ as in diagram \ref{figa2pol1}(a),
and from the propagation of the
boundary state $b_{-n-1,n+1}$ as in diagram \ref{figa2pol3}(a).
The degeneration in energy between these states can be observed
in figures~\ref{spectrum1} and \ref{spectrum2} which show the
spectrum at $B=\frac{1}{n}$ with $n=1$ or $n=5$ respectively.
The degeneration is displayed even more clearly by an animated plot
of the spectrum on this paper's web page at
{\tt http://www.mth.kcl.ac.uk/\~{}delius/reflection.html} which
plots the spin 2 conserved charge of the states against their
energy.

\subsection{$a_3^{(1)}$ ATFT\label{a3}}
For purely accidental reasons we have studied $a_4^{(1)}$ Toda
theory in detail instead of $a_3^{(1)}$.
However $a_3^{(1)}$ has a self-conjugate representation,
as do all $a_n^{(1)}$ Toda theories with $n$ odd.
This is a reason why one
might wish to look at $a_3^{(1)}$ Toda theory in detail.
The single stationary soliton
solutions for solitons in the self-conjugate representation
on the half-line have a slightly different form from the others.
Usually the
soliton solutions on the left half line are obtained by placing a mirror
antisoliton on the unphysical right half line. For the self-conjugate
soliton the soliton and its mirror antisoliton collapse into a single
soliton because they are both stationary and of the same type.
One still obtains  real-valued non-singular vacuum
solutions after analytic continuation of these solutions to real coupling.
However one might expect that this difference in the classical theory
might lead to some new phenomenon in the reflection
amplitudes for the boundary conditions corresponding to a self-conjugate
representation.

In the $a_3^{(1)}$ Toda theory the
$(+ - + -)$, $(- + - +)$ and $(- - - -)$ boundary conditions
correspond to a soliton in the self-dual representation
$\Lambda_2$.
We will give some of the reflection amplitudes
for this case. The theory contains boundary states obtained by
binding $n_1$ particles of type $1$ and $n_3$ particles of type
$3$ to the boundary. The reflection amplitudes for particles
of type $1$ and $2$ reflecting off these boundaries are
\begin{align}\label{k312}
K_1^{n_1,n_3}(\th)\ = K_1&(\theta)
\bl 1 -(2n_1+1)\frac B2\br\
\bl 3 +(2n_3+1) \frac B2\br
&(a)\nn \\
&\times\bl 1 -(2n_1-1)\frac B2\br\
\bl 3 +(2n_3-1) \frac B2\br
&(b)\\
&\times\bl 1 +(2n_1-1)\frac B2\br\
\bl 3 -(2n_3-1) \frac B2\br
&(c)\nn \\
&\times\bl-1 -(2n_3+1) \frac B2\br
\bl-3 -(2n_1+1)\frac B2\br\
&(d)\nn
\end{align}
\begin{align}\label{k322}
K_2^{n_1,n_3}(\th)\ = K_2&(\theta)
\bl 2 -(2n_1-1)\frac B2\br\
\bl 2 +(2n_3-1) \frac B2\br
&(a)\nn \\
&\times\bl 2 -(2n_3-1) \frac B2\br
\bl 2 +(2n_1-1)\frac B2\br\
&(b)\\
&\times
\bl -(2n_1+1)\frac B2\br\
\bl -4 +(2n_3+1) \frac B2\br
&(c)\nn \\
&\times\bl -(2n_3+1)\frac B2\br\
\bl-4 +(2n_1+1) \frac B2\br
&(d)\nn
\end{align}
where
\begin{align}
K_1(\th)=&\bl -1+\frac B2\br\;\bl -3\br\;
\bl 4-\frac B2\br\;,\\
K_2(\th)=&\bl -1+\frac B2\br\;\bl -3\br\;
\bl 4-\frac B2\br\;
\bl 1\br\;\bl 3-\frac B2\br\;
\bl -2\br\;\bl-2+\frac B2\br\;.
\end{align}
The reflection amplitudes for particle $3$ are obtained by charge
conjugation,
\begin{equation}
K_3^{n_1,n_3}(\th)=K_1^{n_3,n_1}(\th).
\end{equation}
The pole in $K_1^{n_1,n_3}$ on line (a) is due to the propagation
of boundary state $b_{n_1+1,n_3}$ in the direct channel
(similar to diagram \ref{figa2pol1}(a)), that on line
(b) is due to the propagation of boundary state $b_{n_1-1,n_2}$ in
the crossed channel
(similar to diagram \ref{figa2pol1}(b)).
The pole on line $(c)$ is due to a diagram
similar to that in figure \ref{figa2pol2}.
Also for the poles in $K_2^{n_1,n_3}$
from the first block on line (a) and the first block on line (b)
one easily finds similar diagrams. This explains all poles for
$n_1$ and $n_2$ positive.

However if the pole in $K_1^{n_1,n_3}$ in line (b) is to be interpreted
as being due to the propagation of a state $b_{n_1-1,n_3}$ this means
that there will also have to be states with negative $n_1$ as long as
$n_1+n_2\geq 0$. The corresponding poles in $K_3^{n_1,0}$ also
require states with negative $n_3$. The reflection amplitude
$K_2^{n_1,n_3}$ develops additional poles in the first block
in line (c) if $n_1<0$ and from the first block in line (c) if
$n_3<0$. These poles indicate the existence of more
boundary bound states. We have not performed the bootstrap
calculations to determine the reflection amplitudes off these new
states. Thus it is an open question whether there is a consistent
explanation of all the poles in this case.

\subsection{$a_4^{(1)}$ ATFT}
This case we will treat in detail.
In this subsection we have
\begin{equation}
\bl a \br \equiv \frac{\sin(\frac{\th}{2i} +\frac{a\pi}{10})}
{\sin(\frac{\th}{2i} -\frac{a\pi}{10})}\;.
\end{equation}
The fusion rules for the particles are that particles $a$ and $b$
can fuse to give particle $c$ if either $a+b=c$ or $a+b=c+5$.
The corresponding
fusion angles are
\begin{equation}
\theta_{ab}^c=\left\{\begin{array}{ll}
\frac{\pi}{5}(a+b)&\text{if } a+b=c\\
\frac{\pi}{5}(10-a-b)&\text{if } a+b=c+5
\end{array}\right.
\end{equation}

In the case of $a_4^{(1)}$ there are two classes of inequivalent
solitonic boundary conditions.
We will treat them in turns.

\subsubsection{The first class of boundary conditions}

The first class contains the boundary conditions
$(- + + + -)$, $(+ + + - -)$, $(+ + - - +)$,
$(- - + + +)$ and
$(+ - - + +)$. Their vacuum solutions are obtained by putting
solitons from the first or fourth fundamental representation in
front of the boundary.
The corresponding vacuum reflection amplitudes are
\begin{align}
K_1^{(1)0,0}(\th) = K_4^{(1)0,0}(\th)
&= K_1(\th) \bl \frac 52 - \frac B2 \br\, \bl
\frac 52 + \frac B2 \br\, \bl -\frac 12 - \frac B2 \br\, \bl -\frac 92 +
\frac B2 \br\;,\\
K_2^{(1)0,0}(\th) = K_3^{(1)0,0}(\th)
&= K_2(\th) \bl \frac 12 - \frac B2 \br\, \bl
\frac 92 + \frac B2 \br\, \bl \frac 32 - \frac B2 \br\, \bl \frac 72 +
\frac B2 \br\;.
\end{align}
$K_1$ and $K_2$ are given by \eq{Ka} and their poles on the physical
strip were explained in section \ref{fixed}.

The poles in $K_2^{(1)}(\th)$ and $K_3^{(1)}(\th)$ from the block
$\left( \frac 12 - \frac B2 \right)$ come from bound states of particles
$2$ and $3$ with the boundary respectively. Bootstrapping on these,
one discovers that there is a whole lattice of boundary bound
states $b_{n_2,n_3}$, where $n_i$ gives the number of particles of
type $i$ bound to the boundary. We allow $n_2$ or $n_3$ to become
negative, as long as the sum $n_2+n_3$ remains non-negative.
The reflection amplitudes off these
boundary bound states are found to be
\begin{align}
\label{k11}
K_1^{(1)n_2,n_3}(\th)\ =  K_1&(\theta)
 \bl\frac 32 -(2n_2-1)\frac B2\br\
\bl\frac 72 +(2n_3-1) \frac B2\br
&(a)\nn \\
&\times\bl\frac 52 -(2n_3-1)\frac B2\br\
\bl\frac 52 +(2n_2-1) \frac B2\br
&(b)\nn \\
&\times\bl-\frac 12 -(2n_2+1)\frac B2\br\
\bl-\frac 92 +(2n_3+1) \frac B2\br
&(c)\nn \\
&\times\bl-\frac 32 -(2n_3+1)\frac B2\br\
\bl-\frac 72 +(2n_2+1) \frac B2\br
&(d)
\end{align}
\begin{align}\label{k21}
K_2^{(1)n_2,n_3}(\th)\ = K_2(\theta)
&\bl\frac 12 -(2n_2+1)\frac B2\br\
\bl\frac 92 +(2n_3+1) \frac B2\br
&(a)\nn \\
&\times\bl\frac 12 -(2n_2-1)\frac B2\br\
\bl\frac 92 +(2n_3-1) \frac B2\br
&(b)\nn \\
&\times\bl\frac 32 -(2n_3-1)\frac B2\br\
\bl\frac 72 +(2n_2-1) \frac B2\br
&(c)\nn \\
&\times\bl\frac 32 +(2n_2-1)\frac B2\br\
\bl\frac 72 -(2n_3-1) \frac B2\br
&(d)\nn \\
&\times\bl\frac 52 -(2n_2-1)\frac B2\br\
\bl\frac 52 +(2n_3-1) \frac B2\br
&(e)\nn \\
&\times\bl-\frac 12 -(2n_3+1)\frac B2\br\
\bl-\frac 92 +(2n_2+1) \frac B2\br
&(f)\nn \\
&\times\bl-\frac 32 -(2n_2+1)\frac B2\br\
\bl-\frac 72 +(2n_3+1) \frac B2\br
&(g)\nn \\
&\times\bl-\frac 52 -(2n_3+1)\frac B2\br\
\bl-\frac 52 +(2n_2+1) \frac B2\br\;.
&(h)
\end{align}
The reflection amplitudes for particles $3$
and $4$ are obtained by charge conjugation,
\begin{equation}
K_3^{(1)n_2,n_3}(\theta)=K_2^{(1)n_3,n_2}(\theta),~~~~~
K_4^{(1)n_2,n_3}(\theta)=K_1^{(1)n_3,n_2}(\theta)
\end{equation}
and it is therefore sufficient if we explain the pole structure of
the reflection amplitudes for particles $1$ and $2$, the poles in the
reflection amplitudes for particles $3$ and $4$ are then explained by
the charge conjugated diagrams.
If $n_2+n_3$ equals to $0$ or $1$ then some
coincidences occur in the above products
of blocks. We will therefore first concentrate on the case
$n_2+n_3>1$.

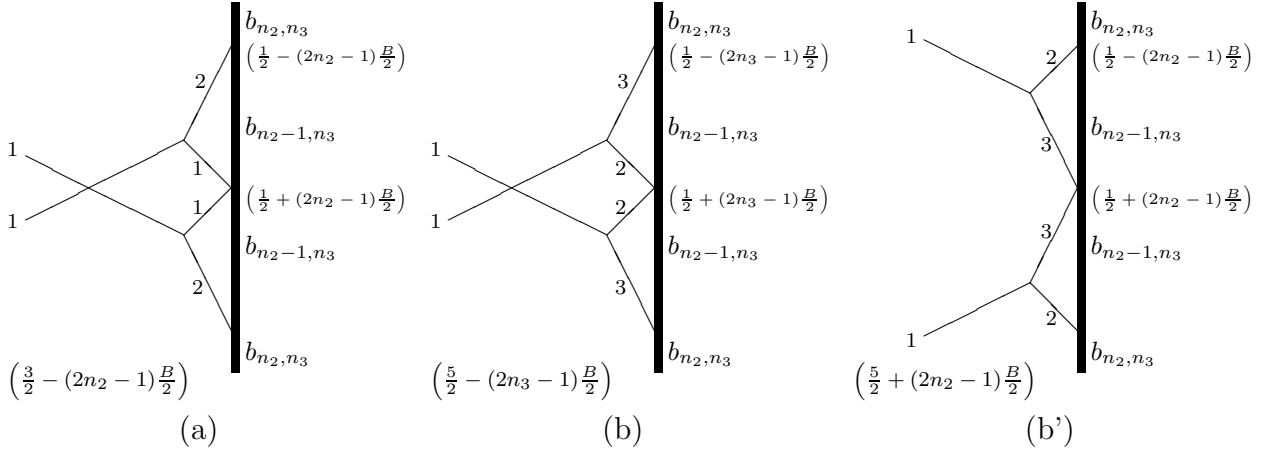
\begin{figure}[p]
\begin{picture}(480,190)(30,-15)
\put(120,10){\rule{3pt}{140pt}}
\put(120,26){\line(-1,2){18}}
\put(120,134){\line(-1,-2){18}}
\put(120,80){\line(-1,-1){18}}
\put(120,80){\line(-1,1){18}}
\put(102,62){\line(-2,1){60}}
\put(102,98){\line(-2,-1){60}}
\put(35,65){\shs{\scs{$1$}}}
\put(35,92){\shs{\scs{$1$}}}
\put(105,40){\shs{\scs{$2$}}}
\put(106,118){\shs{\scs{$2$}}}
\put(105,70){\shs{\scs{$1$}}}
\put(105,85){\shs{\scs{$1$}}}
\put(125,15){\shs{\fns{$b_{n_2,n_3}$}}}
\put(125,55){\shs{\fns{$b_{n_2-1,n_3}$}}}
\put(125,100){\shs{\fns{$b_{n_2-1,n_3}$}}}
\put(125,140){\shs{\fns{$b_{n_2,n_3}$}}}
\put(125,128){\shs{\tiny{$\(\frac 12-(2n_2-1)\frac B2\)$}}}
\put(125,74){\shs{\tiny{$\(\frac 12+(2n_2-1)\frac B2\)$}}}
%
%
\put(280,10){\rule{3pt}{140pt}}
\put(280,26){\line(-1,2){18}}
\put(280,134){\line(-1,-2){18}}
\put(280,80){\line(-1,-1){18}}
\put(280,80){\line(-1,1){18}}
\put(262,62){\line(-2,1){60}}
\put(262,98){\line(-2,-1){60}}
\put(195,65){\shs{\scs{$1$}}}
\put(195,92){\shs{\scs{$1$}}}
\put(265,40){\shs{\scs{$3$}}}
\put(266,118){\shs{\scs{$3$}}}
\put(265,70){\shs{\scs{$2$}}}
\put(265,85){\shs{\scs{$2$}}}
\put(285,15){\shs{\fns{$b_{n_2,n_3}$}}}
\put(285,55){\shs{\fns{$b_{n_2-1,n_3}$}}}
\put(285,100){\shs{\fns{$b_{n_2-1,n_3}$}}}
\put(285,140){\shs{\fns{$b_{n_2,n_3}$}}}
\put(285,128){\shs{\tiny{$\(\frac 12-(2n_3-1)\frac B2\)$}}}
\put(285,74){\shs{\tiny{$\(\frac 12+(2n_3-1)\frac B2\)$}}}
%
%
\put(440,10){\rule{3pt}{140pt}}
\put(440,26){\line(-1,1){18}}
\put(440,134){\line(-1,-1){18}}
\put(440,80){\line(-1,-2){18}}
\put(440,80){\line(-1,2){18}}
\put(422,44){\line(-2,-1){40}}
\put(422,116){\line(-2,1){40}}
\put(375,20){\shs{\scs{$1$}}}
\put(375,135){\shs{\scs{$1$}}}
\put(428,28){\shs{\scs{$2$}}}
\put(428,127){\shs{\scs{$2$}}}
\put(426,61){\shs{\scs{$3$}}}
\put(426,94){\shs{\scs{$3$}}}
\put(445,15){\shs{\fns{$b_{n_2,n_3}$}}}
\put(445,55){\shs{\fns{$b_{n_2-1,n_3}$}}}
\put(445,100){\shs{\fns{$b_{n_2-1,n_3}$}}}
\put(445,140){\shs{\fns{$b_{n_2,n_3}$}}}
\put(445,128){\shs{\tiny{$\(\frac 12-(2n_2-1)\frac B2\)$}}}
\put(445,74){\shs{\tiny{$\(\frac 12+(2n_2-1)\frac B2\)$}}}
\put(35,5){\shs{\scs{$\left(\frac 32 -(2n_2-1)\frac B2\right)$}}}
\put(195,5){\shs{\scs{$\left(\frac 52 -(2n_3-1)\frac B2\right)$}}}
\put(355,5){\shs{\scs{$\left(\frac 52 +(2n_2-1)\frac B2\right)$}}}
\put(100,-15){\shs{ (a)}}
\put(260,-15){\shs{ (b)}}
\put(420,-15){\shs{ (b')}}
\end{picture}
\caption{\it Poles in $K_1^{(1)n_2,n_3}$ with $n_2+n_3>0$.}\label{figa4pol1}
\end{figure}

$K_1^{(1)n_2,n_3}$ has physical strip poles in lines (a) and (b)
which are explained by the first two diagrams in figure
\ref{figa4pol1}.
If $n_3\leq 0$ the pole from the first factor on line (b)
is not on the physical sheet and diagram \ref{figa4pol1}(b) does not
exist.
If $n_2\leq 0$ then the second factor on line (b) is on the
physical sheet and is explained by the third diagram in figure
\ref{figa4pol1}.
It can be checked that the reflections in the middle of the three
processes in figure~\ref{figa4pol1} all occur at a zero in the
corresponding reflection amplitudes, and therefore these diagrams yield
simple poles as expected.

\begin{figure}[p]
\begin{picture}(480,190)(20,-15)
\put(120,10){\rule{3pt}{140pt}}
\put(120,55){\line(-1,-2){15}}
\put(120,105){\line(-1,2){20}}
\put(95,25){\shs{\scs{$2$}}}
\put(95,150){\shs{\scs{$2$}}}
\put(125,30){\shs{\fns{$b_{n_2,n_3}$}}}
\put(125,77){\shs{\fns{$b_{n_2+1,n_3}$}}}
\put(125,125){\shs{\fns{$b_{n_2,n_3}$}}}
%
%
\put(280,10){\rule{3pt}{140pt}}
\put(280,110){\line(-1,-2){40}}
\put(280,50){\line(-1,2){47}}
\put(230,25){\shs{\scs{$2$}}}
\put(230,150){\shs{\scs{$2$}}}
\put(285,30){\shs{\fns{$b_{n_2,n_3}$}}}
\put(285,77){\shs{\fns{$b_{n_2-1,n_3}$}}}
\put(285,125){\shs{\fns{$b_{n_2,n_3}$}}}
%
%
\put(440,10){\rule{3pt}{140pt}}
\put(440,26){\line(-1,2){18}}
\put(440,134){\line(-1,-2){18}}
\put(440,80){\line(-1,-1){18}}
\put(440,80){\line(-1,1){18}}
\put(422,62){\line(-2,1){60}}
\put(422,98){\line(-2,-1){60}}
\put(355,65){\shs{\scs{$2$}}}
\put(355,92){\shs{\scs{$2$}}}
\put(425,40){\shs{\scs{$3$}}}
\put(426,118){\shs{\scs{$3$}}}
\put(425,70){\shs{\scs{$1$}}}
\put(425,85){\shs{\scs{$1$}}}
\put(445,15){\shs{\fns{$b_{n_2,n_3}$}}}
\put(445,55){\shs{\fns{$b_{n_2,n_3-1}$}}}
\put(445,100){\shs{\fns{$b_{n_2,n_3-1}$}}}
\put(445,140){\shs{\fns{$b_{n_2,n_3}$}}}
\put(445,128){\shs{\tiny{$\(\frac 12-(2n_3-1)\frac B2\)$}}}
\put(445,74){\shs{\tiny{$\(\frac 32+(2n_3-1)\frac B2\)$}}}
%
\put(35,5){\shs{\scs{$\left(\frac 12 -(2n_2+1)\frac B2\right)$}}}
\put(195,5){\shs{\scs{$\left(\frac 12 -(2n_2-1)\frac B2\right)$}}}
\put(355,5){\shs{\scs{$\left(\frac 32 -(2n_3-1)\frac B2\right)$}}}
\put(100,-15){\shs{ (a)}}
\put(260,-15){\shs{ (b)}}
\put(420,-15){\shs{ (c)}}
\end{picture}
\caption{\it Poles in $K_2^{(1)n_2,n_3}$ for $n_2+n_3>0$.}\label{figa4pol2}
\end{figure}
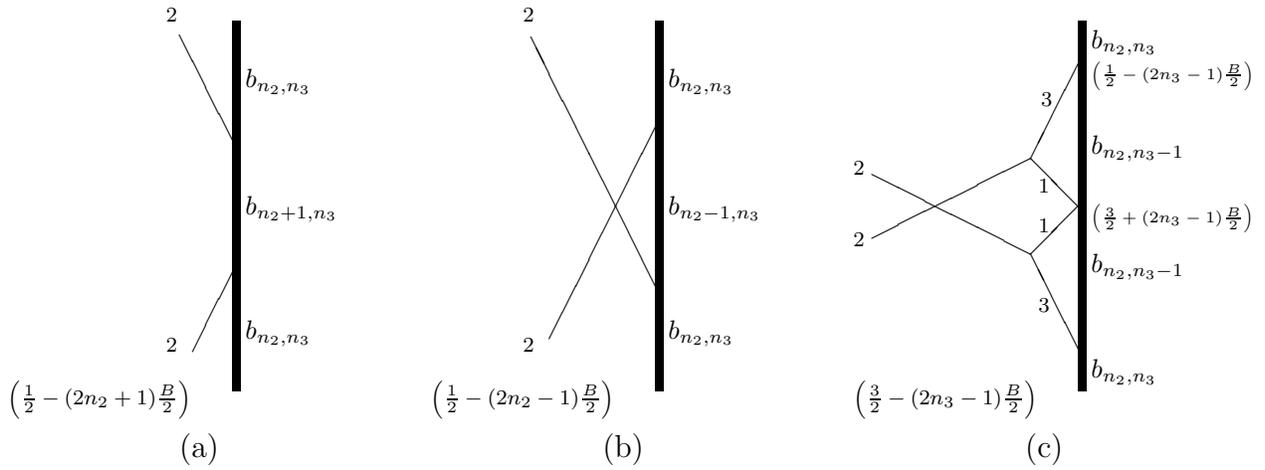

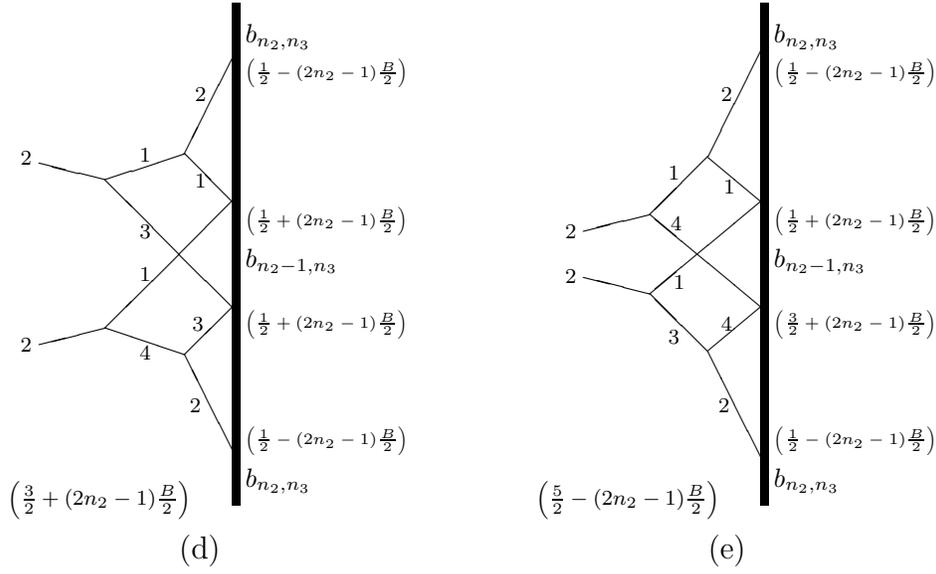
\begin{figure}[p]
\begin{picture}(480,240)(0,-15)
\put(120,5){\rule{3pt}{190pt}}
\put(120,26){\line(-1,2){18}}
\put(120,174){\line(-1,-2){18}}
\put(120,80){\line(-1,-1){18}}
\put(120,120){\line(-1,1){18}}
\put(120,80){\line(-1,1){48}}
\put(120,120){\line(-1,-1){48}}
\put(102,62){\line(-3,1){30}}
\put(102,138){\line(-3,-1){30}}
\put(72,72){\line(-4,-1){25}}
\put(72,128){\line(-4,1){25}}
\put(40,63){\shs{\scs{$2$}}}
\put(40,134){\shs{\scs{$2$}}}
\put(104,40){\shs{\scs{$2$}}}
\put(106,158){\shs{\scs{$2$}}}
\put(85,60){\shs{\scs{$4$}}}
\put(85,135){\shs{\scs{$1$}}}
\put(85,90){\shs{\scs{$1$}}}
\put(85,106){\shs{\scs{$3$}}}
\put(105,71){\shs{\scs{$3$}}}
\put(106,125){\shs{\scs{$1$}}}
\put(125,12){\shs{\fns{$b_{n_2,n_3}$}}}
\put(125,95){\shs{\fns{$b_{n_2-1,n_3}$}}}
\put(125,180){\shs{\fns{$b_{n_2,n_3}$}}}
\put(125,167){\shs{\tiny{$\(\frac 12-(2n_2-1)\frac B2\)$}}}
\put(125,111){\shs{\tiny{$\(\frac 12+(2n_2-1)\frac B2\)$}}}
\put(125,72){\shs{\tiny{$\(\frac 12+(2n_2-1)\frac B2\)$}}}
\put(125,28){\shs{\tiny{$\(\frac 12-(2n_2-1)\frac B2\)$}}}
%
\put(320,5){\rule{3pt}{190pt}}
\put(320,23){\line(-1,2){20}}
\put(320,177){\line(-1,-2){20}}
\put(320,80){\line(-6,-5){20}}
\put(320,120){\line(-6,5){20}}
\put(320,80){\line(-6,5){42}}
\put(320,120){\line(-6,-5){42}}
\put(300,63){\line(-1,1){22}}
\put(300,137){\line(-1,-1){22}}
\put(278,85){\line(-4,1){25}}
\put(278,115){\line(-4,-1){25}}
%
%
\put(246,89){\shs{\scs{$2$}}}
\put(246,106){\shs{\scs{$2$}}}
\put(304,40){\shs{\scs{$2$}}}
\put(305,158){\shs{\scs{$2$}}}
\put(285,66){\shs{\scs{$3$}}}
\put(285,128){\shs{\scs{$1$}}}
\put(287,87){\shs{\scs{$1$}}}
\put(286,109){\shs{\scs{$4$}}}
\put(305,71){\shs{\scs{$4$}}}
\put(306,123){\shs{\scs{$1$}}}
\put(325,12){\shs{\fns{$b_{n_2,n_3}$}}}
\put(325,95){\shs{\fns{$b_{n_2-1,n_3}$}}}
\put(325,180){\shs{\fns{$b_{n_2,n_3}$}}}
\put(325,167){\shs{\tiny{$\(\frac 12-(2n_2-1)\frac B2\)$}}}
\put(325,111){\shs{\tiny{$\(\frac 12+(2n_2-1)\frac B2\)$}}}
\put(325,72){\shs{\tiny{$\(\frac 32+(2n_2-1)\frac B2\)$}}}
\put(325,28){\shs{\tiny{$\(\frac 12-(2n_2-1)\frac B2\)$}}}
\put(35,5){\shs{\scs{$\left(\frac 32 + (2n_2-1)\frac B2\right)$}}}
\put(235,5){\shs{\scs{$\left(\frac 52 - (2n_2-1)\frac B2 \right)$}}}
\put(100,-15){\shs{ (d)}}
\put(300,-15){\shs{ (e)}}
\end{picture}
\caption{\it Poles in $K_2^{(1)n_2,n_3}$ if
$n_2>0$.}\label{figa4pol3}
\end{figure}

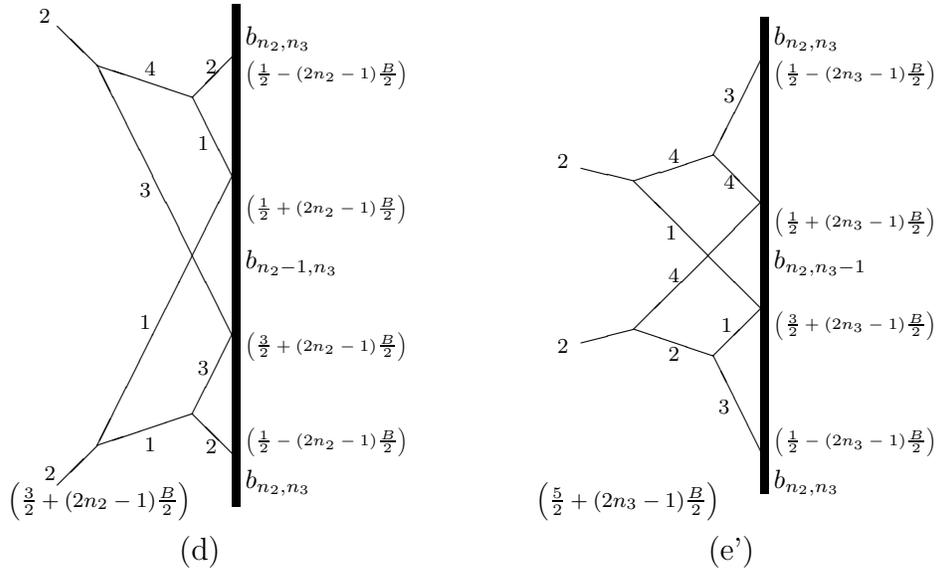
\begin{figure}[p]
\begin{picture}(480,240)(0,-15)
\put(120,5){\rule{3pt}{190pt}}
\put(120,25){\line(-1,1){15}}
\put(120,175){\line(-1,-1){15}}
\put(120,70){\line(-1,-2){15}}
\put(120,130){\line(-1,2){15}}
\put(120,70){\line(-1,2){51}}
\put(120,130){\line(-1,-2){51}}
\put(105,40){\line(-3,-1){36}}
\put(105,160){\line(-3,1){36}}
\put(69,28){\line(-1,-1){15}}
\put(69,172){\line(-1,1){15}}
\put(49,15){\shs{\scs{$2$}}}
\put(47,188){\shs{\scs{$2$}}}
\put(110,25){\shs{\scs{$2$}}}
\put(110,169){\shs{\scs{$2$}}}
\put(87,26){\shs{\scs{$1$}}}
\put(87,168){\shs{\scs{$4$}}}
\put(85,72){\shs{\scs{$1$}}}
\put(85,122){\shs{\scs{$3$}}}
\put(107,55){\shs{\scs{$3$}}}
\put(107,140){\shs{\scs{$1$}}}
\put(125,12){\shs{\fns{$b_{n_2,n_3}$}}}
\put(125,95){\shs{\fns{$b_{n_2-1,n_3}$}}}
\put(125,180){\shs{\fns{$b_{n_2,n_3}$}}}
\put(125,167){\shs{\tiny{$\(\frac 12-(2n_2-1)\frac B2\)$}}}
\put(125,116){\shs{\tiny{$\(\frac 12+(2n_2-1)\frac B2\)$}}}
\put(125,64){\shs{\tiny{$\(\frac 32+(2n_2-1)\frac B2\)$}}}
\put(125,28){\shs{\tiny{$\(\frac 12-(2n_2-1)\frac B2\)$}}}
%
%
\put(320,10){\rule{3pt}{180pt}}
\put(320,26){\line(-1,2){18}}
\put(320,174){\line(-1,-2){18}}
\put(320,80){\line(-1,-1){18}}
\put(320,120){\line(-1,1){18}}
\put(320,80){\line(-1,1){48}}
\put(320,120){\line(-1,-1){48}}
\put(302,62){\line(-3,1){30}}
\put(302,138){\line(-3,-1){30}}
\put(272,72){\line(-4,-1){20}}
\put(272,128){\line(-4,1){20}}
\put(243,63){\shs{\scs{$2$}}}
\put(243,133){\shs{\scs{$2$}}}
\put(304,40){\shs{\scs{$3$}}}
\put(306,158){\shs{\scs{$3$}}}
\put(285,60){\shs{\scs{$2$}}}
\put(285,135){\shs{\scs{$4$}}}
\put(285,90){\shs{\scs{$4$}}}
\put(284,106){\shs{\scs{$1$}}}
\put(305,71){\shs{\scs{$1$}}}
\put(306,125){\shs{\scs{$4$}}}
\put(325,12){\shs{\fns{$b_{n_2,n_3}$}}}
\put(325,95){\shs{\fns{$b_{n_2,n_3-1}$}}}
\put(325,180){\shs{\fns{$b_{n_2,n_3}$}}}
\put(325,167){\shs{\tiny{$\(\frac 12-(2n_3-1)\frac B2\)$}}}
\put(325,111){\shs{\tiny{$\(\frac 12+(2n_3-1)\frac B2\)$}}}
\put(325,72){\shs{\tiny{$\(\frac 32+(2n_3-1)\frac B2\)$}}}
\put(325,28){\shs{\tiny{$\(\frac 12-(2n_3-1)\frac B2\)$}}}
\put(35,5){\shs{\scs{$\left(\frac 32 + (2n_2-1)\frac B2\right)$}}}
\put(235,5){\shs{\scs{$\left(\frac 52 + (2n_3-1)\frac B2 \right)$}}}
\put(100,-15){\shs{ (d)}}
\put(300,-15){\shs{ (e')}}
\end{picture}
\caption{\it Pole (d) in $K_2^{(1)n_2,n_3}$ if $n_2\leq0$
and pole (e) if
$n_3<0$.}\label{figa4pol4}
\end{figure}

The physical strip poles in $K_2^{(1)n_2,n_3}$ can be explained by the
diagrams in figures \ref{figa4pol2}, \ref{figa4pol3} and
\ref{figa4pol4}.
We have drawn diagram \ref{figa4pol3}(d) for the case $n_2>0$. If
$n_2\leq 0$ then the diagram changes slightly to \ref{figa4pol4}(d).
Diagram \ref{figa4pol3}(e) only exists if $n_2>0$. For $n_2\leq 0$ the
corresponding pole moves off the physical strip.
There is an additional pole on the physical strip in the case of
$n_3<0$ in the second block in line (e) of $K_2^{(1)n_2,n_3}$.
This pole can be explained in terms of diagram \ref{figa4pol4}(e).

The diagrams \ref{figa4pol3} and
\ref{figa4pol4} all have
$13$ internal lines and $5$ loops. This would lead to a third
order poles. However it can be checked that the two reflection
factors appearing in every diagram contribute a zero, thereby
reducing the diagram to a simple pole. The scattering amplitudes
which appear in these diagrams do contribute neither a pole nor a
zero. Note that the diagrams \ref{figa4pol3}(e) and \ref{figa4pol4}(e) are
not as symmetric as we have drawn, rather the two reflections are taking
place at different angles, as indicated by the numbers in parentheses.

Next we turn to the case of $n_2=-n_3$ which is rather special.
In $K_1^{(1)n_2,n_3}$ there is a cancellation between line
(a) and line (d), leaving
\begin{align}
K_1^{(1)-n,n}(\th)\ =\ &\bl\frac 52 -(2n+1) \frac B2\br
\bl\frac 52 -(2n-1)\frac B2\br\
&(b)\\
&\times\bl-\frac 12 +(2n-1)\frac B2\br\
\bl-\frac 92 +(2n+1) \frac B2\br
&(c)\nn
\end{align}
In this case none of the diagrams in figure~\ref{figa4pol1}
exist, because they would all involve a state with $n_2+n_3<0$ which
we postulated not to exist. The poles in line (b) from
$\bl \frac 52 - (2n+1)\frac B2 \br$ and
$\bl \frac 52 - (2n-1)\frac B2 \br$
correspond to the first two processes in figure~\ref{figa4pol5}.

\begin{figure}[p]
\begin{picture}(380,190)(0,-15)
\put(120,10){\rule{3pt}{140pt}}
\put(120,55){\line(-1,-2){15}}
\put(120,105){\line(-1,2){20}}
\put(95,25){\shs{\scs{$1$}}}
\put(95,150){\shs{\scs{$1$}}}
\put(125,30){\shs{\fns{$b_{-n,n}$}}}
\put(125,77){\shs{\fns{$b_{-n-1,n+1}$}}}
\put(125,125){\shs{\fns{$b_{-n,n}$}}}
%
%
\put(280,10){\rule{3pt}{140pt}}
\put(280,110){\line(-1,-2){42}}
\put(280,50){\line(-1,2){47}}
\put(230,25){\shs{\scs{$1$}}}
\put(230,150){\shs{\scs{$1$}}}
\put(285,30){\shs{\fns{$b_{-n,n}$}}}
\put(285,77){\shs{\fns{$b_{-n+1,n-1}$}}}
\put(285,125){\shs{\fns{$b_{-n,n}$}}}
\put(45,5){\shs{\scs{$\left(\frac 52 - (2n+1)\frac B2\right)$}}}
\put(205,5){\shs{\scs{$\left(\frac 52 - (2n-1)\frac B2 \right)$}}}
\put(110,-15){\shs{(a)}}
\put(270,-15){\shs{(b)}}
\end{picture}
\caption{\it Poles in $K_1^{(1)-n,n}$.} \label{figa4pol5}
\end{figure}
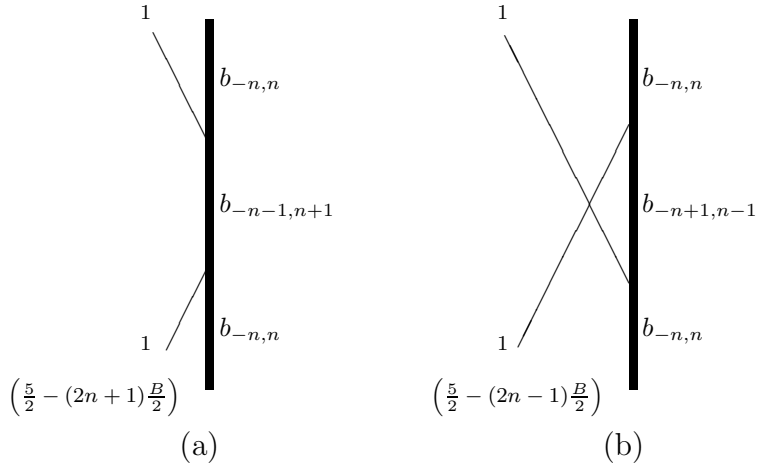


\begin{figure}[p]
\begin{picture}(325.00,150.00)
(0,-15)
\put(120.00,10.00){\rule{3.00\unitlength}{140.00\unitlength}}
\put(125.00,22.00){\shs{\fns{$b_{-n,n}$}}}
\put(125.00,94.00){\shs{\fns{$b_{-n+1,n}$}}}
\put(125.00,135.00){\shs{\fns{$b_{-n,n}$}}}
\put(125.00,39.00){\shs{\scs{$\left(\frac 52 - (2n+1)\frac B2\right)$}}}
\put(125.00,67.00){\shs{\scs{$\left(\frac 12 + (2n_2-1)\frac B2\right)$}}}
\put(320.00,10.00){\rule{3.00\unitlength}{140.00\unitlength}}
\put(320.00,26.00){\line(-1,1){18.00}}
\put(320.00,134.00){\line(-1,-1){18.00}}
\put(320.00,80.00){\line(-1,-2){18.00}}
\put(320.00,80.00){\line(-1,2){18.00}}
\put(302.00,44.00){\line(-2,-1){60.00}}
\put(302.00,116.00){\line(-2,1){60.00}}
\put(235.00,10.00){\shs{\scs{$2$}}}
\put(235.00,143.00){\shs{\scs{$2$}}}
\put(308.00,28.00){\shs{\scs{$4$}}}
\put(308.00,127.00){\shs{\scs{$4$}}}
\put(306.00,61.00){\shs{\scs{$1$}}}
\put(306.00,94.00){\shs{\scs{$1$}}}
\put(325.00,15.00){\shs{\fns{$b_{-n,n}$}}}
\put(325.00,51.00){\shs{\fns{$b_{-n-1,n+1}$}}}
\put(325.00,100.00){\shs{\fns{$b_{-n-1,n+1}$}}}
\put(325.00,140.00){\shs{\fns{$b_{-n,n}$}}}
\put(325.00,128.00){\shs{\scs{$\left(\frac 52 + (2n+1)\frac B2\right)$}}}
\put(325.00,73.00){\shs{\scs{$\left(\frac 12 - (2n+1)\frac B2\right)$}}}
\put(43.00,5.00){\shs{\scs{$\left(\frac 32 - (2n+1)\frac B2\right)$}}}
\put(245.00,5.00){\shs{\scs{$\left(\frac 32 - (2n+1)\frac B2 \right)$}}}
\put(110.00,-15.00){\shs{(a)}}
\put(310.00,-15.00){\shs{(b)}}
\put(120.00,80.00){\line(-1,-1){40.00}}
\put(80.00,40.00){\line(4,1){40.00}}
\put(80.00,40.00){\line(-2,-1){41.00}}
\put(120.00,80.00){\line(-1,1){40.00}}
\put(80.00,120.00){\line(4,-1){40.00}}
\put(80.00,120.00){\line(-2,1){41.00}}
\put(35.00,19.00){\makebox(0,0)[rc]{\shs{\scs{$2$}}}}
\put(35.00,141.00){\makebox(0,0)[rc]{\shs{\scs{$2$}}}}
\put(101.00,41.00){\makebox(0,0)[ct]{\shs{\scs{$1$}}}}
\put(101.00,118.00){\makebox(0,0)[cb]{\shs{\scs{$1$}}}}
\put(97.00,96.00){\makebox(0,0)[rt]{\shs{\scs{$1$}}}}
\put(97.00,65.00){\makebox(0,0)[rb]{\shs{\scs{$1$}}}}
\end{picture}
\caption{\it Extra pole in $K_2^{(1)-n,n}$ for a) $n\geq 0$ and b)
$n<0$.} \label{figa4pol5b}
\end{figure}
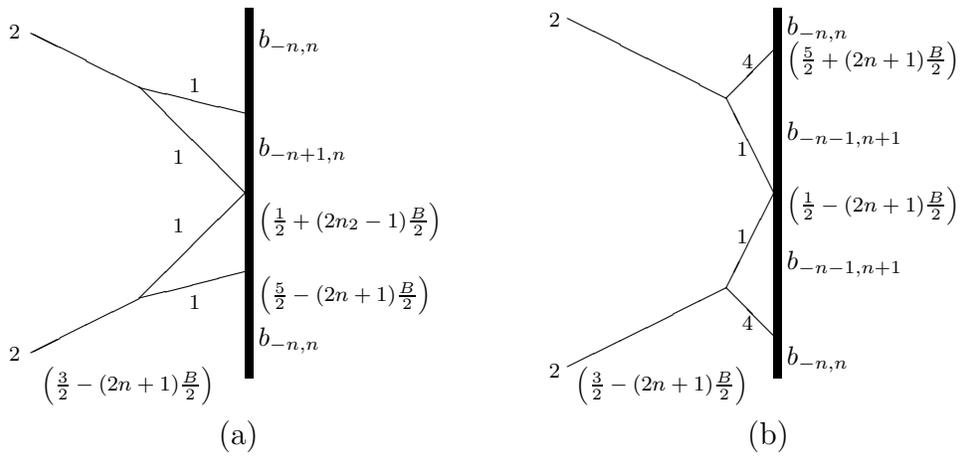


In $K_2^{(1)n_2,n_3}$ with $n_2=-n_3$ there are cancellations
between lines (b) and (f), (c) and (g) and (e) and (h), leaving only
\begin{align}
K_2^{(1)-n,n}(\th)\ =\ &\bl\frac 12 +(2n-1)\frac B2\br\
\bl\frac 92 +(2n+1) \frac B2\br
&(a)\nn \\
&\times\bl\frac 32 -(2n+1)\frac B2\br\
\bl\frac 72 -(2n-1) \frac B2\br
&(d)
\end{align}
The pole in line (a) still receives a contribution from the process
in figure~\ref{figa4pol2}(a). However
this pole does not satisfy the bootstrap equation which one would
expect on the basis of  figure~\ref{figa4pol2}(a). The reason is that
there is now a second diagram
contributing to this pole, shown in figure~\ref{figa4pol5b}(a).

The processes in figures \ref{figa4pol2}(b) to
\ref{figa4pol4}(e) disappear, because they would involve boundary states
with $n_2+n_3<0$. The pole in line (d) from $\bl\frac 32 -(2n+1)\frac
B2\br$ is explained by the diagram \ref{figa4pol5b}(d).

For a given  $B$ with $0<B<1$ the following boundary bound
states exist
\bea\label{boundsa4}
&b_{n_2,n_3}\;,&\mbox{for all integers} \hs{3pt} n_2,n_3 \hs{3pt} \mbox{with}
\hs{3pt} n_2+n_3 > 0 \hs{3pt} \mbox{and} \hs{3pt}
-\frac 1{2B} - \frac 12 < n_2,n_3  < \frac 1{2B}+\frac 12\;, \nn \\
&b_{n,-n}  \hs{3pt} \mbox{and} \hs{3pt} b_{-n,n}\;, \hs{4pt}
&\mbox{for all integers} \hs{3pt} n \hs{3pt} \mbox{with}
\hs{3pt} 0\leq n < \frac 5{2B}+\frac 12\;. \nn
\eea

Finally we consider the case where $n_2=-n_3+1$. In
$K_1^{(1)n_2,n_3}$ the two blocks in line (b) coincide and
give rise to a double pole. The reason for this is the fact that the
reflection process in the middle of diagram \ref{figa4pol1}(b) or
\ref{figa4pol1}(c) no longer provides a zero because it is a reflection
of a boundary with $n_2+n_3=0$.
Similarly in $K_2^{(1)n_2,n_3}$ the blocks in line (c) combine
with those in line (d) to form double poles, as do the two blocks in
line (e). Again this is because one of the reflection amplitudes in
each of the diagrams \ref{figa4pol2}(c), \ref{figa4pol3}(a), and
\ref{figa4pol3}(b) or \ref{figa4pol4}(b) ceases to supply a zero.

\subsubsection{The second class of boundary conditions}

The second class contains the boundary conditions $(- + + - +)$,
$(- + - + +)$, $(+ + - + -)$, $(- + - - -)$, $(+ - + + -)$, $(+ -
+ - +)$, $(- - + - -)$, $(- - - + -)$, $(- - - - +)$
and $(+ - - - -)$. Their vacuum solutions are obtained
by putting a soliton from the second or third fundamental
representation in front of the boundary.
The corresponding vacuum reflection amplitudes are
\begin{align}
K_1^{(2)}(\th) = K_4^{(2)}(\th)
= K_1(\th) &\bl \frac 12 - \frac B2 \br\, \bl
\frac 92 + \frac B2 \br\, \bl \frac 32 - \frac B2 \br\, \bl \frac 72 +
\frac B2 \br\;.\\
K_2^{(2)}(\th) = K_3^{(1)}(\th)
= K_2(\th) &\bl \frac 12 - \frac B2 \br\, \bl
\frac 92 + \frac B2 \br\, \bl \frac 32 - \frac B2 \br\, \bl \frac 72 +
\frac B2 \br \nn \\
&\times\bl \frac 52 -\frac B2 \br \, \bl \frac 52 + \frac B2 \br\,
\bl -\frac 12 - \frac B2 \br\;.
\end{align}
All of these reflection amplitudes have a pole from a block
$\bl \frac 12 - \frac B2 \br$, indicating that all particles can
bind to the boundary. Indeed we find a whole lattice of boundary bound
states $b_{n_1,n_2,n_3,n_4}$. For reasons to be explained below
this lattice is restricted from below by the three conditions
\begin{align}
n_1+n_2&\geq 0, &n_2+n_3&\geq 0, &n_3+n_4&\geq 0.
\end{align}
As previously the array is also bounded above because the poles at
which the bound states are created move off the physical strip.
Using the usual bootstrap equations, the reflection amplitudes off these
boundary bound states are found to be
\begin{align}\label{k12}
K_1^{(2)n_1,n_2,n_3,n_4}(\th)\ = K_1^{(1)n_2,n_3}&(\theta)
\bl\frac 12 -(2n_1+1)\frac B2\br\
\bl\frac 92 +(2n_4+1) \frac B2\br
&(e)\nn \\
&\times\bl\frac 12 -(2n_1-1)\frac B2\br\
\bl\frac 92 +(2n_4-1) \frac B2\br
&(f)\\
&\times\bl\frac 32 +(2n_1-1)\frac B2\br\
\bl\frac 72 -(2n_4-1) \frac B2\br
&(g)\nn \\
&\times\bl-\frac 52 -(2n_4+1)\frac B2\br\
\bl-\frac 52 +(2n_1+1) \frac B2\br
&(h)\nn
\end{align}
\begin{align}\label{k22}
K_2^{(2)n_1,n_2,n_3,n_4}(\th)\ = K_2^{(1)n_2,n_3}&(\theta)
\bl\frac 32 -(2n_1-1)\frac B2\br\
\bl\frac 72 +(2n_4-1) \frac B2\br
&(i)\nn \\
&\times\bl\frac 52 -(2n_4-1)\frac B2\br\
\bl\frac 52 +(2n_1-1) \frac B2\br
&(j)\\
&\times\bl-\frac 12 -(2n_1+1)\frac B2\br\
\bl-\frac 92 +(2n_4+1) \frac B2\br
&(k)\nn \\
&\times\bl-\frac 32 -(2n_4+1)\frac B2\br\
\bl-\frac 72 +(2n_1+1) \frac B2\br
&(l)\nn
\end{align}

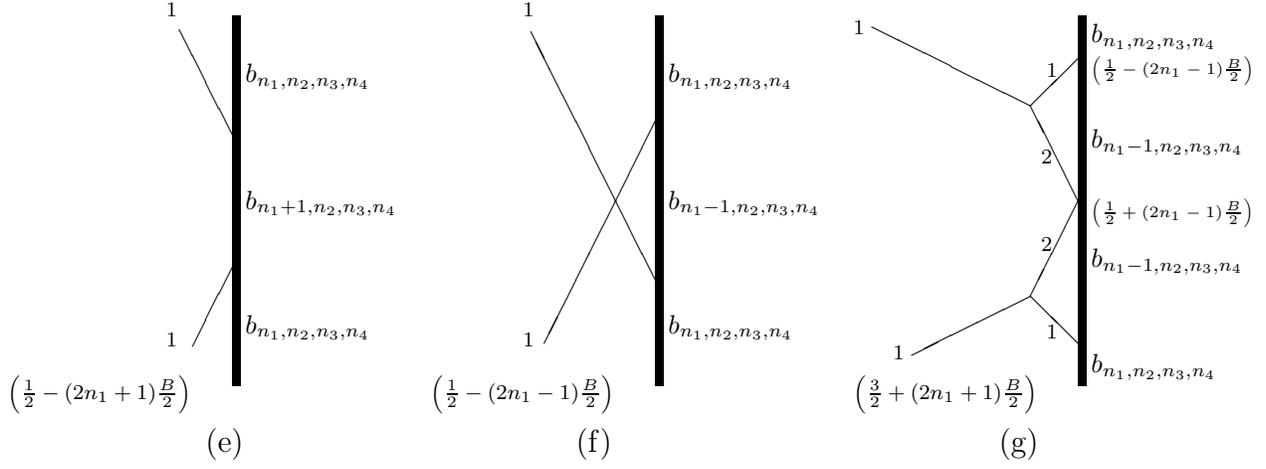
\begin{figure}[p]
\begin{picture}(480,190)(20,-15)
\put(120,10){\rule{3pt}{140pt}}
\put(120,55){\line(-1,-2){15}}
\put(120,105){\line(-1,2){20}}
\put(95,25){\shs{\scs{$1$}}}
\put(95,150){\shs{\scs{$1$}}}
\put(125,30){\shs{\fns{$b_{n_1,n_2,n_3,n_4}$}}}
\put(125,77){\shs{\fns{$b_{n_1+1,n_2,n_3,n_4}$}}}
\put(125,125){\shs{\fns{$b_{n_1,n_2,n_3,n_4}$}}}
%
%
\put(280,10){\rule{3pt}{140pt}}
\put(280,110){\line(-1,-2){42}}
\put(280,50){\line(-1,2){47}}
\put(230,25){\shs{\scs{$1$}}}
\put(230,150){\shs{\scs{$1$}}}
\put(285,30){\shs{\fns{$b_{n_1,n_2,n_3,n_4}$}}}
\put(285,77){\shs{\fns{$b_{n_1-1,n_2,n_3,n_4}$}}}
\put(285,125){\shs{\fns{$b_{n_1,n_2,n_3,n_4}$}}}
%
%
\put(440,10){\rule{3pt}{140pt}}
\put(440,26){\line(-1,1){18}}
\put(440,134){\line(-1,-1){18}}
\put(440,80){\line(-1,-2){18}}
\put(440,80){\line(-1,2){18}}
\put(422,44){\line(-2,-1){45}}
\put(422,116){\line(-2,1){60}}
\put(370,20){\shs{\scs{$1$}}}
\put(355,143){\shs{\scs{$1$}}}
\put(428,28){\shs{\scs{$1$}}}
\put(428,127){\shs{\scs{$1$}}}
\put(426,61){\shs{\scs{$2$}}}
\put(426,94){\shs{\scs{$2$}}}
\put(445,15){\shs{\fns{$b_{n_1,n_2,n_3,n_4}$}}}
\put(445,55){\shs{\fns{$b_{n_1-1,n_2,n_3,n_4}$}}}
\put(445,100){\shs{\fns{$b_{n_1-1,n_2,n_3,n_4}$}}}
\put(445,140){\shs{\fns{$b_{n_1,n_2,n_3,n_4}$}}}
\put(445,128){\shs{\tiny{$\(\frac 12-(2n_1-1)\frac B2\)$}}}
\put(445,74){\shs{\tiny{$\(\frac 12+(2n_1-1)\frac B2\)$}}}
\put(35,5){\shs{\scs{$\left(\frac 12 - (2n_1+1)\frac B2\right)$}}}
\put(195,5){\shs{\scs{$\left(\frac 12 - (2n_1-1)\frac B2 \right)$}}}
\put(355,5){\shs{\scs{$\left(\frac 32+ (2n_1+1)\frac B2 \right)$}}}
\put(110,-15){\shs{(e)}}
\put(250,-15){\shs{(f)}}
\put(410,-15){\shs{(g)}}
\end{picture}
\caption{\it Poles in $K_1^{(2)n_1,n_2,n_3,n_4}$}\label{figa4pol6}
\end{figure}


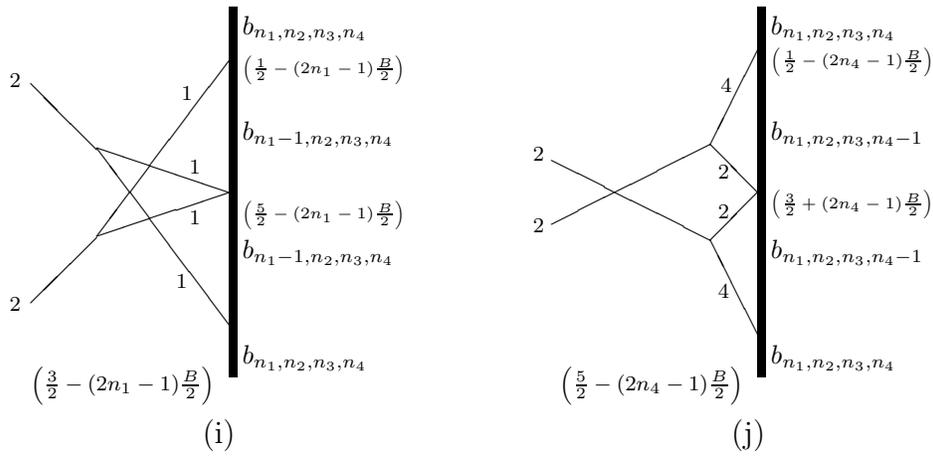
\begin{figure}[p]
\begin{picture}(480,190)(0,-15)
\put(120,10){\rule{3pt}{140pt}}
\put(120,30){\line(-3,4){50}}
\put(120,130){\line(-3,-4){50}}
\put(120,80){\line(-3,-1){50}}
\put(120,80){\line(-3,1){50}}
\put(70,63){\line(-1,-1){25}}
\put(70,96){\line(-1,1){25}}
\put(37,35){\shs{\scs{$2$}}}
\put(37,120){\shs{\scs{$2$}}}
\put(100,44){\shs{\scs{$1$}}}
\put(102,115){\shs{\scs{$1$}}}
\put(105,68){\shs{\scs{$1$}}}
\put(105,87){\shs{\scs{$1$}}}
\put(125,15){\shs{\fns{$b_{n_1,n_2,n_3,n_4}$}}}
\put(125,55){\shs{\fns{$b_{n_1-1,n_2,n_3,n_4}$}}}
\put(125,100){\shs{\fns{$b_{n_1-1,n_2,n_3,n_4}$}}}
\put(125,140){\shs{\fns{$b_{n_1,n_2,n_3,n_4}$}}}
\put(125,70){\shs{\tiny{$\(\frac 52-(2n_1-1)\frac B2\)$}}}
\put(125,125){\shs{\tiny{$\(\frac 12-(2n_1-1)\frac B2\)$}}}
%
\put(320,10){\rule{3pt}{140pt}}
\put(320,26){\line(-1,2){18}}
\put(320,134){\line(-1,-2){18}}
\put(320,80){\line(-1,-1){18}}
\put(320,80){\line(-1,1){18}}
\put(302,62){\line(-2,1){60}}
\put(302,98){\line(-2,-1){60}}
\put(235,65){\shs{\scs{$2$}}}
\put(235,92){\shs{\scs{$2$}}}
\put(305,40){\shs{\scs{$4$}}}
\put(306,118){\shs{\scs{$4$}}}
\put(305,70){\shs{\scs{$2$}}}
\put(305,85){\shs{\scs{$2$}}}
\put(325,15){\shs{\fns{$b_{n_1,n_2,n_3,n_4}$}}}
\put(325,55){\shs{\fns{$b_{n_1,n_2,n_3,n_4-1}$}}}
\put(325,100){\shs{\fns{$b_{n_1,n_2,n_3,n_4-1}$}}}
\put(325,140){\shs{\fns{$b_{n_1,n_2,n_3,n_4}$}}}
\put(325,128){\shs{\tiny{$\(\frac 12-(2n_4-1)\frac B2\)$}}}
\put(325,74){\shs{\tiny{$\(\frac 32+(2n_4-1)\frac B2\)$}}}
%
\put(45,5){\shs{\scs{$\left(\frac 32 - (2n_1-1)\frac B2\right)$}}}
\put(245,5){\shs{\scs{$\left(\frac 52 - (2n_4-1)\frac B2 \right)$}}}
\put(110,-15){\shs{(i)}}
\put(310,-15){\shs{(j)}}
\end{picture}
\caption{\it Poles in $K_2^{(2)n_1,n_2,n_3,n_4}$}\label{figa4pol7}
\end{figure}


\begin{figure}[p]
\begin{picture}(400,190)(0,-15)
\put(120,10){\rule{3pt}{140pt}}
\put(120,26){\line(-1,2){18}}
\put(120,134){\line(-1,-2){18}}
\put(120,80){\line(-1,-1){18}}
\put(120,80){\line(-1,1){18}}
\put(102,62){\line(-2,1){60}}
\put(102,98){\line(-2,-1){60}}
\put(35,65){\shs{\scs{$2$}}}
\put(35,92){\shs{\scs{$2$}}}
\put(105,40){\shs{\scs{$1$}}}
\put(106,118){\shs{\scs{$1$}}}
\put(105,70){\shs{\scs{$4$}}}
\put(105,85){\shs{\scs{$4$}}}
\put(125,15){\shs{\fns{$b_{n_1,n_2,n_3,n_4}$}}}
\put(125,55){\shs{\fns{$b_{n_1-1,n_2,n_3,n_4}$}}}
\put(125,100){\shs{\fns{$b_{n_1-1,n_2,n_3,n_4}$}}}
\put(125,140){\shs{\fns{$b_{n_1,n_2,n_3,n_4}$}}}
\put(125,128){\shs{\tiny{$\(\frac 12-(2n_1-1)\frac B2\)$}}}
\put(125,74){\shs{\tiny{$\(\frac 52+(2n_1-1)\frac B2\)$}}}
%
\put(320,10){\rule{3pt}{140pt}}
\put(320,26){\line(-1,1){18}}
\put(320,134){\line(-1,-1){18}}
\put(320,80){\line(-1,-2){18}}
\put(320,80){\line(-1,2){18}}
\put(302,44){\line(-2,-1){60}}
\put(302,116){\line(-2,1){60}}
\put(235,10){\shs{\scs{$2$}}}
\put(235,143){\shs{\scs{$2$}}}
\put(308,28){\shs{\scs{$1$}}}
\put(308,127){\shs{\scs{$1$}}}
\put(306,61){\shs{\scs{$3$}}}
\put(306,94){\shs{\scs{$3$}}}
\put(325,15){\shs{\fns{$b_{n_1,n_2,n_3,n_4}$}}}
\put(325,55){\shs{\fns{$b_{n_1-1,n_2,n_3,n_4}$}}}
\put(325,100){\shs{\fns{$b_{n_1-1,n_2,n_3,n_4}$}}}
\put(325,140){\shs{\fns{$b_{n_1,n_2,n_3,n_4}$}}}
\put(325,128){\shs{\tiny{$\(\frac 12-(2n_1-1)\frac B2\)$}}}
\put(325,74){\shs{\tiny{$\(\frac 32+(2n_1-1)\frac B2\)$}}}
\put(45,5){\shs{\scs{$\left(\frac 32 - (2n_1-1)\frac B2\right)$}}}
\put(245,5){\shs{\scs{$\left(\frac 52 + (2n_1-1)\frac B2 \right)$}}}
\put(110,-15){\shs{(i)}}
\put(310,-15){\shs{(j)}}
\end{picture}
\caption{\it Poles in $K_2^{(2)n_1,n_2,n_3,n_4}$ with $n_1\leq 0$.}\label{figa4pol8}
\end{figure}
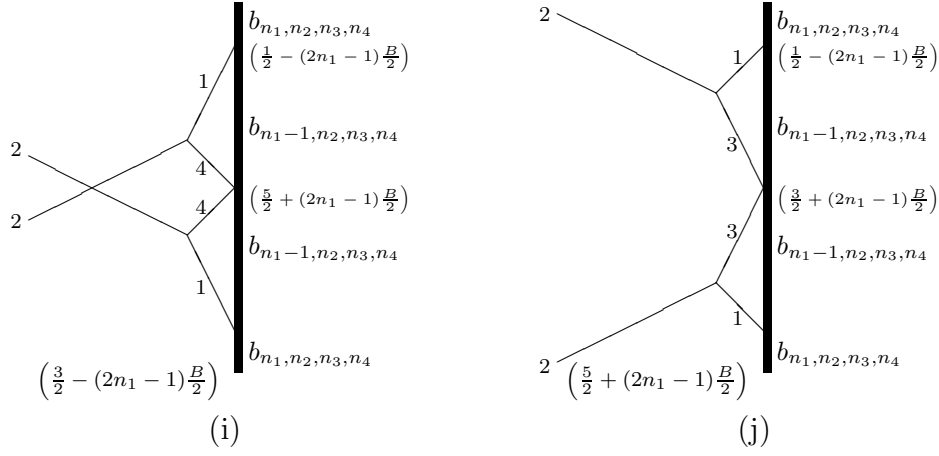


\begin{figure}[p]
\begin{picture}(380,190)(0,-15)
\put(120,10){\rule{3pt}{140pt}}
\put(120,55){\line(-1,-2){15}}
\put(120,105){\line(-1,2){20}}
\put(95,25){\shs{\scs{$1$}}}
\put(95,150){\shs{\scs{$1$}}}
\put(125,30){\shs{\fns{$b_{-n_2,n_2,n_3,n_4}$}}}
\put(125,77){\shs{\fns{$b_{-n_2-1,n_2+1,n_3,n_4}$}}}
\put(125,125){\shs{\fns{$b_{-n_2,n_2,n_3,n_4}$}}}
%
%
\put(280,10){\rule{3pt}{140pt}}
\put(280,110){\line(-1,-2){42}}
\put(280,50){\line(-1,2){47}}
\put(230,25){\shs{\scs{$1$}}}
\put(230,150){\shs{\scs{$1$}}}
\put(285,30){\shs{\fns{$b_{-n_2,n_2,n_3,n_4}$}}}
\put(285,77){\shs{\fns{$b_{-n_2+1,n_2-1,n_3,n_4}$}}}
\put(285,125){\shs{\fns{$b_{-n_2,n_2,n_3,n_4}$}}}
\put(45,5){\shs{\scs{$\left(\frac 32 - (2n_2+1)\frac B2\right)$}}}
\put(205,5){\shs{\scs{$\left(\frac 32 - (2n_2-1)\frac B2 \right)$}}}
\put(110,-15){\shs{(g)}}
\put(270,-15){\shs{(a)}}
\end{picture}
\caption{\it Poles in $K_1^{(2)n_1,n_2,n_3,n_4}$ if $n_1=-n_2$.}\label{figa4pol9}
\end{figure}
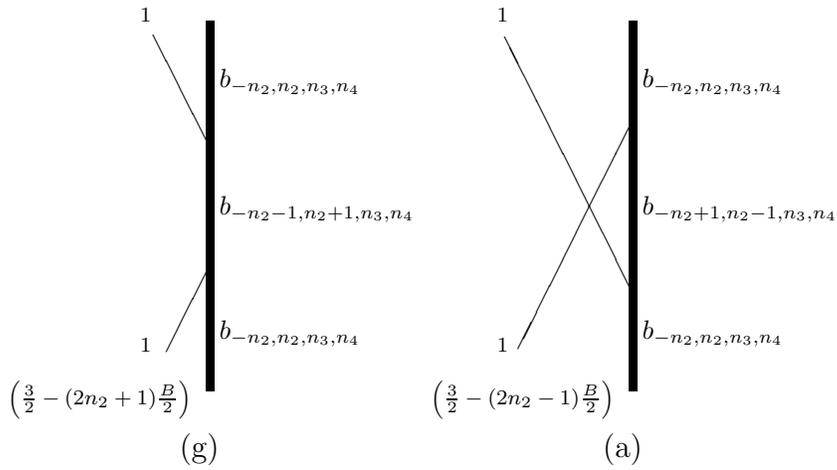


\begin{figure}
\begin{picture}(480,190)(0,-15)
\put(120,10){\rule{3pt}{140pt}}
\put(120,26){\line(-1,1){18}}
\put(120,134){\line(-1,-1){18}}
\put(120,80){\line(-1,-2){18}}
\put(120,80){\line(-1,2){18}}
\put(102,44){\line(-2,-1){60}}
\put(102,116){\line(-2,1){60}}
\put(35,10){\shs{\scs{$2$}}}
\put(35,145){\shs{\scs{$2$}}}
\put(108,28){\shs{\scs{$1$}}}
\put(108,127){\shs{\scs{$1$}}}
\put(106,61){\shs{\scs{$3$}}}
\put(106,94){\shs{\scs{$3$}}}
\put(125,15){\shs{\fns{$b_{-n_2,n_2,n_3,n_4}$}}}
\put(125,55){\shs{\fns{$b_{-n_2+1,n_2-1,n_3,n_4}$}}}
\put(125,140){\shs{\fns{$b_{-n_2,n_2,n_3,n_4}$}}}
\put(125,128){\shs{\tiny{$\(\frac 32-(2n_2-1)\frac B2\)$}}}
\put(125,74){\shs{\tiny{$\(\frac 12+(2n_2-1)\frac B2\)$}}}
%
\put(320,10){\rule{3pt}{140pt}}
\put(320,26){\line(-1,2){18}}
\put(320,134){\line(-1,-2){18}}
\put(320,80){\line(-1,-1){18}}
\put(320,80){\line(-1,1){18}}
\put(302,62){\line(-2,1){60}}
\put(302,98){\line(-2,-1){60}}
\put(235,65){\shs{\scs{$2$}}}
\put(235,92){\shs{\scs{$2$}}}
\put(305,40){\shs{\scs{$1$}}}
\put(306,118){\shs{\scs{$1$}}}
\put(305,70){\shs{\scs{$4$}}}
\put(305,85){\shs{\scs{$4$}}}
\put(325,15){\shs{\fns{$b_{-n_2,n_2,n_3,n_4}$}}}
\put(325,55){\shs{\fns{$b_{-n_2+1,n_2-1,n_3,n_4}$}}}
\put(325,140){\shs{\fns{$b_{n_1,n_2,n_3,n_4}$}}}
\put(325,128){\shs{\tiny{$\(\frac 32-(2n_2-1)\frac B2\)$}}}
\put(325,74){\shs{\tiny{$\(\frac 32+(2n_2-1)\frac B2\)$}}}
\put(45,5){\shs{\scs{$\left(\frac 32 + (2n_2-1)\frac B2\right)$}}}
\put(245,5){\shs{\scs{$\left(\frac 52 - (2n_2-1)\frac B2 \right)$}}}
\put(110,-15){\shs{(d)}}
\put(310,-15){\shs{(e)}}
\end{picture}
\caption{\it Poles in $K_2^{(2)n_1,n_2,n_3,n_4}$ if $n_1=-n_2$.}\label{figa4pol10}
\end{figure}
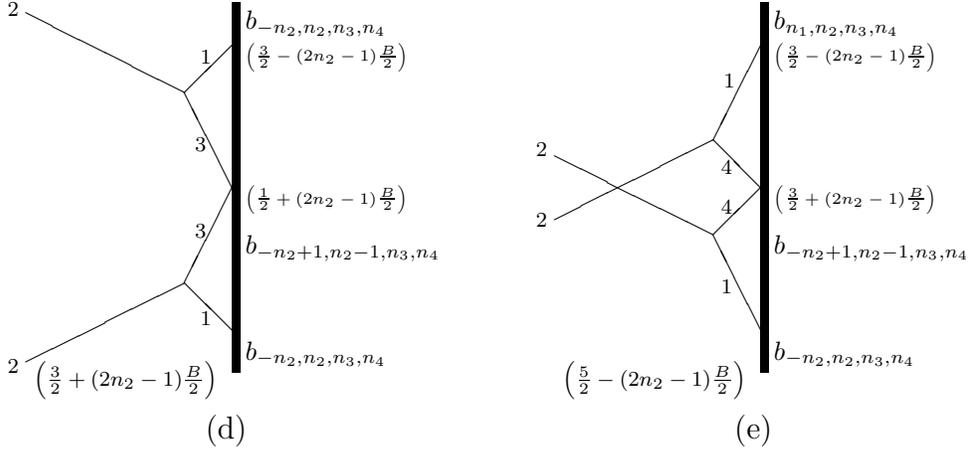

Again the reflection amplitudes for particles $3$ and $4$ are
obtained by charge conjugation. Notice that these reflection
amplitudes for the second boundary condition contain those for the
first boundary condition as a factor. We only need to explain the
poles introduced by the new additional factors. These all depend on
$n_1$ or $n_4$ and must thus be due to diagrams involving the
emission or absorption of particles $1$ or $4$ from the boundary.
For $K_1^{(2)n_1,n_2,n_3,n_4}(\th)$ the relevant diagrams are
displayed in figure~\ref{figa4pol6}, while for
$K_2^{(2)n_1,n_2,n_3,n_4}(\th)$ they are given in figure~\ref{figa4pol7}.

Diagram \ref{figa4pol7}(j) exists only if $n_4>0$. Otherwise the
corresponding pole lies off the physical sheet.
If $n_1\leq0$ then diagram \ref{figa4pol7}(i) changes to diagram
\ref{figa4pol8}(i). Also if $n_1\leq 0$ then the second pole in line
(j) is on the physical sheet, and corresponds to diagram
\ref{figa4pol8}(j).

It is a fun exercise to go through all the special
cases for which blocks in the reflection amplitudes coincide. For
example if $n_4+n_1=0$ then line (i) cancels
against line (l). Correspondingly figures \ref{figa4pol7}(i) and
\ref{figa4pol8}(i)  no longer
provide poles because the reflection amplitudes in the middle of
those diagrams have double zeros. The disappearance of the zero in
line (l) at $n_4+n_1=0$ in turn has the effect that diagrams
\ref{figa4pol7}(j) and \ref{figa4pol8}(j) provide double poles if
$n_4+n_1=1$ which is correct because then the two blocks on line (j)
coincide.

If $n_1+n_2=0$ then the pole on line (f) in $K_1^{(2)n_1,n_2,n_3,n_4}$
disappears and thus diagram \ref{figa4pol6}(f) also has to
disappear. That means that we need to exclude any boundary bound
states with
$n_1+n_2<0$. From the charge conjugated amplitudes we see that
we also have to exclude states with $n_4+n_3<0$. This is the
justification for the restriction on the lattice of states
mentioned above.

The poles in line (g) and (a) in $K_1^{(2)n_1,n_2,n_3,n_4}$ have a new
explanation if $n_1+n_2=0$ in terms of the diagrams in figure
\ref{figa4pol9}, and diagram \ref{figa4pol1}(c) and its corresponding
pole disappear.

At $n_1+n_2=0$ also the diagrams \ref{figa4pol2}(b), \ref{figa4pol3}(d),
\ref{figa4pol4}(d), \ref{figa4pol3}(e), \ref{figa4pol7}(i)
\ref{figa4pol8}(i) and \ref{figa4pol8}(j) disappear. Most of the
corresponding poles also
disappear but the first poles on lines (d) and (e) do not and are
explained by the two new diagrams in figure~\ref{figa4pol10}.

We should still draw more diagrams to explain new poles which
become relevant if one of the indices becomes sufficiently
negative. However we imagine that the reader will be exhausted by
now and so are we and we will therefore stop here. Clearly what is
needed is a more systematic treatment which could explain why
everything falls into place so miraculously.

\section{Discussion}
The most striking result of this paper is probably the fact that the
boundary of real coupling affine Toda theory has such a rich spectrum
of boundary states. Already in the simplest case of $a_2^{(1)}$
Toda theory the spectrum is rather intricate, as displayed in
figures \ref{spectrum1} and \ref{spectrum2}. The pole structure of
the reflection amplitudes necessitates the existence of these
states. If one of them were absent there would be unexplained
poles. However we can not exclude the possibility that there might
be even more states. There are many simple poles in the reflection
amplitudes which we were able to explain by anomalous threshold
diagrams and for which we therefore did not have to introduce
new boundary states. However it is possible
that some of the anomalous threshold poles mask genuine bound
state poles. This does happen in the S-matrices of non self-dual
affine Toda theories \cite{deliu92,corri93}.
Whether it happens in $a_n^{(1)}$ Toda theory with a boundary
could only be decided if one were
able to calculate the residues of the anomalous threshold poles
and compare them to the actual residues of the poles in the
reflection amplitudes to see if there is a mismatch. Such calculations
were carried out in perturbation theory for the Toda S-matrices
\cite{brade91,deliu92} but we do not know how to perform such calculations
in the presence of the boundary.

Another result of this paper, which was not obvious from the
classical analysis, is
that many classically different integrable boundary conditions are
equivalent at the quantum level, in the sense that they give rise
to the same boundary states and the same reflection amplitudes.

When the classical real-valued vacuum solutions of
$a_n^{(1)}$ Toda theory with solitonic boundary conditions
were obtained in
\cite{deliu98} it was remarked that they contain a
continuous parameter. We were wondering what consequence this
vacuum degeneracy would have in the quantum theory. For the
$(- -\cdots- -)$ boundary condition of $a_{2n+1}^{(1)}$ Toda
theory
this degeneracy had been observed earlier in \cite{fujii95}.
In that paper it was implied that the vacuum degeneracy implies
that the theory is ``unstable''. We disagree with that
interpretation. The present paper shows that it is consistent
to assume that the quantum theory picks a unique stable vacuum state.

We have derived reflection amplitudes for those integrable
boundary conditions which satisfy the constraint that the product
of all boundary parameters is $+1$. These are the boundary
conditions for which non-singular real vacuum solutions can be
obtained by analytic continuation of stationary soliton solutions.
We have no results for those boundary conditions for which the
product of all boundary parameters is $-1$.

The Durham group had made a conjecture in \cite{corri94} for the
boundary conditions which has all $C_i=-1$. For $n$ odd this is
a solitonic boundary condition of type $(n+1)/2$
and we can compare their conjecture
with our result. Not surprisingly our results differ because the
calculations in \cite{corri94} were based on a constant vacuum,
whereas we use a vacuum obtained by continuing a stationary soliton
solution, which has the same energy as the constant solution.
So these theories appear to have two different vacuum sectors.
For the $(++\cdots++)$ boundary condition on the other hand the
constant solution is the lowest energy solution and
the conjecture made in \cite{corri94} does agree with
our result \eqref{Ka}.

The results of this paper are of course not rigorous. At several
stages we had to make assumptions. We will list the most important
ones here so that the reader can decide how much confidence he
wants to put into them.

1) We have assumed that the scalar factor
of the soliton reflection matrix is given by \eq{aa} which was the
minimal solution of the equations \eqref{aeq}. However there is
the freedom of including CDD factors of the form \eqref{cdd}. We
believe that inclusion of any of these CDD factors would lead to
an inconsistent pole structure after bootstrap, but we have not
performed a careful analysis of this. The semiclassical
comparision between or reflection matrix and the classical time
delay which we have performed does not detect the CDD factors.

2) We have assumed that the classical solutions to the solitonic boundary
conditions found in \cite{deliu98} are indeed the solutions of
lowest energy. The calculations performed during the work on
\cite{deliu98} have given us strong confidence that this is indeed
the case but we don't know of any way to prove it.

3) The classical vacuum solutions which we use are obtained by
analytic continuation to real coupling of solutions describing a
stationary soliton in front of the boundary in the imaginary
coupling theory. We have assumed that therefore also the particle
reflection amplitudes in the real coupling theory are obtained by
the same analytic continuation of the corresponding breather
reflection amplitude in the imaginary coupling theory. This
assumption can really only be justified a posteriori by the fact
that it has led to consistent results.

Clearly, we have not exhausted the subject of Toda reflection
amplitudes in this paper. While we have indeed given the
reflection amplitudes for all known vacuum solutions satisfying
integrable boundary conditions, we have completely worked out the
reflection
amplitudes for the excited boundary states only for two theories.

We have not discussed the special coincidences and
truncations which can occur at special rational values of the
coupling constant. We have given an example of such a phenomenon
at the end of section \ref{subsect:mixed2} where a coupling constant
$b=\frac{1}{2n}$ with $n$ any integer leads to a degeneracy in the
spectrum.

Furthermore, we have restricted our attention to the regime where
the coupling constant satisfies $B<1$. The spectrum of bound
states changes drastically as soon as $B$ equals $1$.
The study of the case $B>1$ is related to the question of
weak--strong duality. Affine Toda theory in the bulk displays a
fascinating weak--strong coupling duality under which
$\beta\leftrightarrow\frac{4\pi}{\beta}$ or $B\leftrightarrow
2-B$. In particular the S-matrices of $a_n^{(1)}$ Toda theory
are self-dual under
this transformation of the coupling constant.
Now that we have the reflection amplitudes we can study the
interesting question of how S-duality works in the presence of
boundaries. The only thing which we can say so far is that the
Neumann boundary condition is dual to the $(++\cdots++)$ boundary
condition. Previously this duality had been observed in the
sine-Gordon model \cite{corri97} and the
$a_2^{(1)}$ theory \cite{gande98b}.

Without having to do any detailed comparisons we can immediately
say that none of the reflection amplitudes which we have
calculated coincide with the reflection amplitudes which had been
conjectured in \cite{fring94,kim95b,kim,sasak93}. These papers had assumed that the
reflection amplitudes should be invariant under $B\rightarrow 2-B$
which ours manifestly are not.

Working out the explanation of the pole structure of the
reflection amplitudes in the examples of $a_2^{(1)}$ and
$a_4^{(1)}$ Toda theory has been rather tedious.
However it has been quite fascinating to see the
miraculous way in which
these Toda theories have managed to be consistent.
But miracles demand to be studied and (eventually) understood.
This probably will require more cases for more different algebras
to be worked out. It should be possible to give explicit formulas
for the particle reflection factors off the excited boundary states
in $a_n^{(1)}$ theories for
general $n$, starting from our formula \eqref{Kmixed}
for the particle
reflection amplitudes off the vacuum boundary state.

To extend the results to algebras other than $a_n^{(1)}$ one will
have to repeat the analysis of section \ref{sect:an} to find the
soliton reflection matrices as solutions of the corresponding
reflection equations
and then the breather reflection amplitudes
by bootstrap. To formulate the reflection equations one will need
to use the soliton S-matrices which are known for many algebras
\cite{gande96}. We suspect that again the simplest diagonal
solutions of the reflection equations will correspond to the
uniform $(++\cdots ++)$
boundary condition.
Because in all studied cases it has been found that
the particle S-matrices are obtained from the breather S-matrices
by analytical continuation to real coupling, it is to be expected
that the same will hold for the reflection
amplitudes.

To find the reflection amplitudes for other integrable boundary
conditions we used our knowledge of the corresponding vacuum
solutions from \cite{deliu98}. The results of that paper can also
be used to find vacuum solutions for the solitonic boundary conditions
in $c_n^{(1)}$ Toda theories
and therefore the analysis of this paper could be carried over to
those cases.
Clearly one would like to eventually go beyond a case-by-case
analysis and obtain general formulas for the reflection
amplitudes. Would formulas as elegant as those provided by Dorey
\cite{dor} for the S-matrices in terms of root systems be possible?

We hope that the exact results for the reflection amplitudes,
which we have given in this paper, will be a useful
guide in the attempts to generalise standard quantum field theoretic
techniques to field theories with boundaries.
For example, perturbation theory in the presence of a non-constant
boundary background is a non-trivial matter.
So far only one-loop calculations for constant backgrounds
have been performed \cite{kim,perki98} and our
results confirm them.

\acknowledgements

We would like to thank Peter Bowcock, Edward Corrigan, Patrick Dorey
and Gerard Watts for interesting discussions.

GMG has been supported by EPSRC research grant no. GR/K 79437 and
GWD by an EPSRC advanced fellowship. This collaboration has been
supported in part by EU contract FMRX-CT96-0012.


\parskip 1pt
{\footnotesize
\begin{thebibliography}{99}
%
\bibitem{Arin} A.E. Arinshtein, V.A. Fateev and A.B.
Zamolodchikov, {\it Quantum S-matrix of the 1+1 dimensional Toda
chain}, Phys. Lett. {\bf B87} (1979) 389.
%
\bibitem{ber} D. Bernard and A. LeClair, {\em Quantum Group
Symmetries and Non-Local Currents in 2D QFT}, Commun. Math. Phys.
142, (1991) 99-138.
%
\bibitem{bowco95}
P.\ Bowcock, E.\ Corrigan, P.E.\ Dorey and R.H.\
Rietdijk, {\em Classically integrable boundary conditions for affine
Toda field theories}, Nucl.Phys.\ {\bf B445} (1995), 469; {\tt
hep-th/9501098}.
%
%
\bibitem{bowco96b} P.\ Bowcock, {\em Classical backgrounds and
scattering for affine Toda theory on a half-line}, JHEP 05(1998)008;
{\tt hep-th/9609233}
%
\bibitem{brade90} H.W.\ Braden, E.\ Corrigan, P.E.\ Dorey and R.\
Sasaki, {\em Affine Toda Field Theory and Exact $S$-matrices},
Nucl. Phys.\ {\bf B338} (1990) 689.
%
\bibitem{brade91}
H.W.\ Braden, E.\ Corrigan, P.E.\ Dorey and R.\
Sasaki,
{\em Multiple poles and other features of affine Toda field
theory}, Nucl. Phys.\ {\bf B356} (1991), 469;
%
\bibitem{chere84} I.V.\ Cherednik, {\em Factorizing particles on a
half-line and root systems}, Theor.Math.Phys.\ {\bf 61} (1984)
977.
%
\bibitem{chri} P. Christe and G. Mussardo, {\it Elastic S-matrices
in (1+1) dimensions and Toda field theories}, Int. J. Mod. Phys.
{\bf A5} (1990) 4581.
%
\bibitem{cole} S. Coleman and H.J. Thun, {\em On the Prosaic
Origin of the Double Poles in the Sine-Gordon S-Matrix},
Commun. Math. Phys. {\bf 61} (1978) 31-39.
%
\bibitem{corri93} E. Corrigan, P.E. Dorey and R. Sasaki, {\em
On a generalised bootstrap principle}, Nucl. Phys. {\bf B408}
(1993) 579; {\tt hep-th/9304065}.
%
\bibitem{corri94} E.\ Corrigan, P.E.\ Dorey, R.H.\ Rietdijk and R.\
Sasaki, {\em Affine Toda field theory on a half-line},
Phys.Lett.\ {\bf B333} (1994) 83; {\tt hep-th/9404108}.
%
\bibitem{corri94b} E.\ Corrigan, P.E.\ Dorey and R.H.\ Rietdijk,
{\em Aspects of affine Toda field theory on a half-line},
Prog.Theor.Phys.Suppl.\ {\bf 118} (1995) 143; {\tt
hep-th/9407148}.
%
\bibitem{corri96} E.\ Corrigan, {\em Integrable Field Theory with
Boundary Conditions}, in {\it Frontiers in Quantum Field Theory},
eds. Chao-Zheng Zha and Ke Wu, (World Scientific 1998) pp9--32,
{\tt hep-th/9612138}
%
\bibitem{corri97} E.\ Corrigan, {\em On duality and reflection factors
for the sinh-Gordon model}, Int. J. Mod. Phys. {\bf A13} (1998)
2709--2722, {\tt hep-th/9707235}.
%
\bibitem{deliu92} G. W. Delius, M. T. Grisaru and D. Zanon,
{\em Exact S-Matrices for nonsimply-laced affine Toda theories},
Nucl. Phys. {\bf B382} (1992) 365-408;
{\tt hep-th/9201067}.
%
\bibitem{deliu98} G.W.\ Delius, {\em Restricting affine Toda theory to
the half--line}, J. High Energy Phys. 9809 (1998) 016;
{\tt hep-th/9807189}.
%
\bibitem{deliu98b} G.W.\ Delius, {\em Soliton-preserving boundary
conditions in affine Toda field theories},
Phys. Lett. {\bf B444} (1998) 217-223; {\tt hep-th/9809140}.
%
\bibitem{deveg93} H.J.\ de Vega and A.\ Gonz\'{a}lez Ruiz, {\em
Boundary $K$-matrices for the six vertex and the $n(2n-1)$ $A_{n-1}$
vertex models}, Jour.Phys.\ {\bf A26} (1993), L519
%
\bibitem{dor} P.E.\ Dorey, {\em Root systems and purely elastic
S-matrices, I \& II}, Nucl. Phys. B358 (1991) 654, and
Nucl. Phys. B374 (1992) 741,
{tt hep-th/9110058}.
%
\bibitem{dorey98} P.E.\ Dorey, R.\ Tateo and G. Watts, {\em
Generalisations of the Coleman--Thun mechanism and boundary reflection
factors}, preprint DTP-98/71, KCL-MTH/98-40, T-98/106, {\tt
hep-th/9810098}.
%
\bibitem{fring94} A.\ Fring and R.\ K{\"o}berle, {\em Factorized
Scattering in the Presence of Reflecting Boundaries}, Nucl.Phys.\ {\bf
B421} (1994), 159; {\tt hep-th/9304141}.\\
A.\ Fring and R.\ K{\"o}berle,
{\em Affine Toda Field Theory in the
Presence of Reflecting Boundaries}, Nucl.Phys.\ {\bf B419} (1994),
647; {\tt hep-th/9309142}.\\
A.\ Fring and R.\ K{\"o}berle, {\em Boundary Bound States in
Affine Toda Field Theory}, Int. J. Mod. Phys. A10 (1995) 739;
{\tt hep-th/9404188}.
%
\bibitem{fujii95} A.\ Fujii and R.\ Sasaki, {\em Boundary Effects in
Integrable Field Theory on a Half-Line}, Prog.Theor.Phys.\ {\bf 93}
(1995), 1123;  {\tt hep-th/9311027}
%
\bibitem{gande95} G.M.\ Gandenberger, {\em Exact $S$-matrices for bound
states of $a_2^{(1)}$ affine Toda solitons}, Nucl.Phys.\ {\bf B449}
(1995), 375; {\tt hep-th/9501136}
%
%
\bibitem{gande96} G.M.\ Gandenberger, {\em Exact $S$-matrices for
Quantum Affine Toda Solitons and their Bound States}, Ph.D.\ thesis,
University of Cambridge 1996, unpublished;\\
{\bf Available as postscript file at:
{\em http://www.damtp.cam.ac.uk/user/hep/publications.html}}
%
\bibitem{gande98} G.M.\ Gandenberger, {\em Trigonometric $S$-matrices,
Affine Toda Solitons and Supersymmetry}, Int.Jour.Mod.Phys. {\bf A13}
(1998), 4553; {\tt hep-th/9703158}
%
\bibitem{gande98b} G.M.\ Gandenberger, {\em On $\ato$ Reflection
Matrices and Affine Toda Theories},
Nucl.Phys. {\bf B542} (1999) 659-693; {\tt hep-th/9806003}
%
\bibitem{gande99} G.M.\ Gandenberger, {\em New non-diagonal solutions
to the $\ao$ boundary Yang-Baxter equation}, in preparation
%
\bibitem{ghosh94} S.\ Ghoshal and A.\ Zamolodchikov, {\em Boundary
$S$-Matrix and Boundary State in Two-Dimensional Integrable Field
Theory}, Int.Jour.Mod.Phys.\ {\bf A9} (1994), 3841; {\tt
hep-th/9306002}
%
\bibitem{ghosh94b} S.\ Ghoshal, {\em Bound State Boundary $S$-Matrix
of the Sine-Gordon Model}, Int. J. Mod. Phys.\ {\bf A9} (1994), 4801;
{\tt hep-th/9310188}
%
\bibitem{hol} T.J. Hollowood, {\em
Quantum Solitons in Affine Toda Field Theories},
{\tt hep-th/9110010}.\\
T.J. Hollowood, {\em Quantizing SL(N) Solitons and the
Hecke Algebra}, Int. J. Mod. Phys. A8 (1993) 947-982; {\tt
hep-th/9203076}.
%
\bibitem{kim95b} J.D.\ Kim, {\em Boundary Reflection Matrix
for $ade$ Affine Toda Theory}, preprint DTP/95-31; {\tt
hep-th/9506031}.
%
\bibitem{kim}
J.D.\ Kim and Y.\ Yoon, {\em Root Systems and Boundary
Bootstrap}, preprint KAIST/THP-96/701; {\tt hep-th/9603111}.\\
J.D. Kim and H. S. Cho, {\em Boundary Reflection Matrix
for $D_4^{(1)}$ Affine Toda Field Theory}, {\tt
hep-th/9505138}.\\
J. D. Kim, {\em Boundary Reflection Matrix in Perturbative
Quantum Field Theory}, Phys. Lett. {\bf B353} (1995) 213;
{\tt hep-th/9504018}.
%
\bibitem{oli} D.I. Olive, M.V. Saveliev and J.W.R. Underwood,
{\em on a solitonic specialisation for the general solutions
of some two-dimensional completely integrable systems},
Phys. Lett. B311 (1993) 117; {\tt hep-th/9212123}.
%
\bibitem{perki98} M.\ Perkins and P.\ Bowcock, {\em Quantum
corrections to the classical reflection factor in $\ato$ Toda field
theory}, preprint DTP-98/49, {\tt hep-th/9807146}.
%
\bibitem{sasak93} R.\ Sasaki, {\em Reflection Bootstrap Equations for
Toda Field Theory}, Hangzhou Proceedings, Interface between Mathematics
and Physics, eds. Werner Nahm and Jian-min Shen
(World Scientific 1994) pp 201--212.; {\tt hep-th/9311027}.
%
%
\bibitem{skori95} S.\ Skorik and H.\ Saleur, {\em
Boundary bound states and boundary bootstrap in the sine-Gordon
model  with Dirichlet boundary conditions},
J.Phys. A28 (1995) 6605, {\tt hep-th/9502011}.
%
\bibitem{Zam} A.B. Zamolodchikov and Al.B. Zamolodchikov,
{em Factorized S-Matrices in Two Dimensions as the Exact Solutions
of Certain Relativistic Quantum Field Theory Models},
Ann. Phys. 120 (1979) 253-291.
%

\end{thebibliography}
}
\end{document}